\documentclass[pdftex, twocolumn,epjc3]{svjour3a}          
\RequirePackage[T1]{fontenc}

\RequirePackage{graphicx}
\graphicspath{{figures/}}
\RequirePackage{mathptmx}      
\RequirePackage{flushend}
\RequirePackage[numbers,sort&compress]{natbib}
\usepackage{subfigure}
\newcommand{\orcid}[1]{\href{https://orcid.org/#1}{\includegraphics[width=10pt]{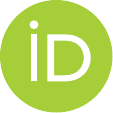}}}

\usepackage{booktabs,multirow}
\usepackage{hhline}
\usepackage{slashed}
\usepackage{amsmath, amssymb}
\usepackage[dvipsnames]{xcolor,colortbl}

\usepackage{xcolor}\colorlet{linkequation}{Blue}
\usepackage[colorlinks=true, 
linkcolor=Red,   
filecolor=Red,   
citecolor=Green, 
urlcolor=Blue,
hyperfootnotes=true]{hyperref}

\newcommand*{\SavedEqref}{}
\let\SavedEqref\eqref
\renewcommand*{\eqref}[1]{%
	\begingroup
	\hypersetup{
		linkcolor=linkequation,
		linkbordercolor=linkequation,
	}%
	\SavedEqref{#1}%
	\endgroup
}
\journalname{Eur. Phys. J. C}

\begin{document}
\title{Probing active-sterile neutrino transition magnetic moment on coherent elastic solar neutrino-nucleus scattering}
\author{M.~Demirci \thanksref{addr1,e1}  \orcid{0000-0003-2504-6251}  \and M.~F.~Mustamin \thanksref{addr1,e2}    \orcid{0000-0003-3996-4651}
}                     
%
\thankstext{e1}{e-mail: mehmetdemirci@ktu.edu.tr (corresponding author)}
\thankstext{e2}{e-mail: mfmustamin@ktu.edu.tr}

\institute{Department of Physics,
	Karadeniz Technical University, Trabzon 61080, Türkiye \label{addr1}} 
\date{Received: date / Revised version: date}
\date{\today}

\titlerunning{Probing active-sterile neutrino transition magnetic moment on CE$\nu$NS}

\maketitle

\begin{abstract}
In the presence of a transition magnetic moment between active and sterile neutrinos, sterile neutrinos could be produced by neutrino beams electromagnetically upscattering on nuclei. We study the active-sterile neutrino transition magnetic moment through this upscattering in the coherent elastic neutrino-nucleus scattering process induced by solar neutrinos. We place new limits on the transition magnetic moment-sterile neutrino mass plane using the latest data from the CDEX-10 experiment. We also provide projected sensitivities for future measurements. 
We observe that the projected sensitivities could cover some regions of the parameter space which were previously unexplored for the sterile neutrino mass up to $\sim$10 MeV.
\end{abstract}
%

%
\section{Introduction}
\label{intro}
Coherent elastic neutrino-nucleus scattering (CE$\nu$NS) is a neutral current process induced by the exchange of a vector $Z$ boson of Standard Model (SM) \cite{Freedman:1973yd}. In this process, a relatively low-energy neutrino interacts with a nucleus as a whole and, as a result of the interaction, the nucleus acquires a kinetic recoil energy that can be measured.  
The first observation of the process was recently carried out by the COHERENT experiment \cite{Akimov:2017ade} with a CsI[Na] scintillating crystal detector using (anti-)neutrinos from pion-decays-at-rest ($\pi$DAR). It was also detected later using a single-phase liquid argon \cite{Akimov:2020pdx} and with a larger sample of CsI[Na] detector \cite{COHERENT:2021xmm}.
Since its first detection, there has been a significant increase in scientific activities, both experimentally and theoretically, related to this process. This is because CE$\nu$NS provides a promising way to probe the SM parameters as well as physics beyond the SM (BSM) at low momentum transfer.
In particular, it has been widely used in examining the weak mixing angle \cite{Canas2018,Cadeddu2019} as well as searching for dark-matter (DM) \cite{Harnik:2012ni,Ge2018,Boehm:2020ltd,Schwemberger:2022fjl,Mishra:2023jlq}, non-standard neutrino interactions (NSI) \cite{Lindner2017,Liao2017,Giunti:2019xpr,Mustamin:2021mtq,Chatterjee:2023}, the effective generalized interactions \cite{Dent:2016wcr,Barranco:2011wx,AristizabalSierra:2018eqm,Flores:2021kzl}, light mediators \cite{Farzan:2018gtr,Cadeddu:2020nbr,Demirci:2021zci,AtzoriCorona:2022moj,Demirci:2024,DeRomeri:2024dbv}, Migdal effect \cite{Herrera:2023xun}, and the electromagnetic properties of neutrinos \cite{Giunti:2015,Studenikin2020,Giunti2023} such as magnetic moment, charge radius, electric charge/millicharge, dipole electric moment and anapole moment \cite{AtzoriCorona:2022,Khan:2023,Khan:2023b}. Moreover, it gives useful information on the nuclear structure, particularly on the nucleus neutron density distribution that is still unknown for most nuclei \cite{Co:2020gwl,Coloma:2020nhf}.

The discovery of neutrino flavor oscillations \cite{Super-Kamiokande:1998kpq,SNO:2001kpb,SNO:2002tuh} implies that neutrino masses are nonzero; this is a fact not taken into account in the SM.
The standard theory must be extended to account for this observation accordingly. 
Many known mechanisms to give neutrinos mass involve fermionic neutral SM-gauge-group singlets, referred to as sterile neutrinos \cite{Pontecorvo:1967fh,Kusenko:2009up,Dasgupta:2021ies}, also so-called heavy neutral leptons.  
Sterile neutrino scenarios are often motivated by solving anomalies or problems, found in short-baseline oscillation and reactor experiments such as MiniBoone \cite{MiniBooNE:2010idf}, MicroBoone \cite{Arguelles:2021meu} and LSND \cite{LSND:1997vun}. A sterile neutrino, which could explain these anomalies, is in the eV-mass range that potentially has implications on nucleosynthesis in core-collapse supernovae \cite{McLaughlin:1999pd,Xiong:2019nvw}. In the higher mass ranges, sterile neutrino could also be a DM candidate \cite{Dodelson:1993je}. 
Apart from these, in many other studies the idea of sterile neutrinos was widely invoked, such as effective neutrino magnetic moment \cite{Balantekin2014}, connection to extra dimensions \cite{Khan:2022bcl}, and the evolution of the Early Universe \cite{Mirizzi:2012we}. 

Future CE$\nu$NS experiments aim to detect extremely low nuclear recoil thresholds from the neutrino-nucleus scattering events. These next-generation activities will not only improve the precision test of CE$\nu$NS in the SM but also will further constrain or give a clue to new physics beyond SM.
One such new physics scenario is the scattering of light active neutrinos into heavy sterile neutrinos through a transition magnetic moment. This process is known as Primakoff upscattering \cite{Domokos:1996cn,Gninenko:1998nn}, originally proposed for neutral-meson photoproduction in a nuclear electric field \cite{Primakoff:1951iae}.
A large transition magnetic moment will raise the possibility of producing the sterile neutrino by electromagnetically upscattering neutrino beams on nuclei \cite{McKeen2010}. To place the limits on the transition magnetic moment, such a dipole portal has been extensively studied using various experimental data from accelerator neutrino sources \cite{Magill:2018jla,Blanco2020,Schwetz2021}, neutral current $\nu_\mu$-nucleus interactions in detector facilities \cite{Gninenko2011}, spallation neutron sources \cite{DeRomeri:2022twg}, nuclear power reactors \cite{Bolton2022}, forward neutrino detection \cite{Ismail:2021dyp}, atmospheric neutrinos \cite{Coloma:2017ppo,Plestid:2020vqf,Atkinson:2021rnp}, direct detection (DD) of DM \cite{Miranda2021,Li2022}, and other cosmological neutrino sources \cite{Brdar:2020quo}. 
Apart from these, it has also been studied in neutrino telescopes \cite{Huang:2022pce}, particle colliders \cite{Antusch:2016ejd}, and diffuse supernova neutrino background \cite{Balantekin:2023jlg}.

In the present work, we explore the active-sterile neutrino transition magnetic moment through CE$\nu$NS with solar neutrinos. In DD experiments solar neutrinos can induce CE$\nu$NS events, as well as elastic neutrino-electron scattering. In this context, we consider the latest data of the CDEX-10 experiment \cite{CDEX:2022mlp} which has a primary goal of searching light DM candidates \cite{CDEX:2018lau}.  The CDEX-10 collaboration has reported neutrino-nucleus event rates from solar neutrinos with $205.4$ kg$\cdot$day exposure using a p-type point contact germanium (PPCGe) detector array.  Accordingly, we derive new constraints on the parameter space of active-sterile neutrino transition magnetic moment vs sterile neutrino mass. We also provide projected sensitivities for next-generation and future scenarios constructed by taking into account experimental developments. Furthermore, we make a comparison of our results with the available limits derived from previous experiments mentioned above, together with other experiments such as XENON \cite{XENON:2020rca}, DUNE \cite{DUNE:2020ypp}, BOREXINO \cite{Borexino:2007kvk}, TEXONO \cite{TEXONO:2009knm}, DRESDEN \cite{Colaresi:2022obx}, NUCLEUS \cite{NUCLEUS:2019kxv}, CHARM \cite{CHARM:1983ayi}, DONUT \cite{DONUT:2001zvi}, NOMAD \cite{NOMAD:1997pcg}, ALEPH \cite{ALEPH:1991qhf}, IceCube \cite{IceCube:2015vkp}, the proposed CENNS \cite{COHERENT:2019kwz}, FLArE \cite{Batell:2021blf} and SHiP \cite{Alekhin:2015byh} at CERN, as well as future DD experiment of DARWIN \cite{DARWIN:2016hyl}.

The remainder of this paper is organized as follows. In Sect.~\ref{sec:cevns}, we briefly introduce the theoretical expressions of the CE$\nu$NS in both SM and in the presence of active-sterile neutrino transition magnetic moment. Details of the data analysis are presented in Sect.~\ref{section:data}. Then, in Sect.~\ref{sec:results}, we illustrate the predicted event rates and the statistical analysis results. Finally, we provide concluding remarks in Sect.~\ref{sec:conc}.
\section{Theoretical Framework}\label{sec:cevns}
\subsection{CE$\nu$NS in the SM}
In the CE$\nu$NS process, neutrinos can undergo coherent elastic scattering simultaneously with all nucleons of the nucleus through neutral-current interaction with only a small amount of the momentum transferred to the nucleus. 
At the tree level in the SM, the differential cross-section of the neutrino scattering off the nucleus is given by 
\begin{eqnarray}
	\begin{split}
		\biggl[	\frac{d\sigma (E_\nu, T_{nr})}{dT_{nr}} \biggr]^{\text{CE}\nu\text{NS}}_\text{SM}=&\frac{G_F^2 m_\mathcal{N}}{\pi} \Big(Q^\text{SM}_V\Big)^2 F^2(|\vec{q}|^2) \\ &\times\Big(1-\frac{m_\mathcal{N}T_{nr}}{2E_\nu^2}\Big),
	\end{split}
\end{eqnarray}
where $G_F$ is the Fermi constant, $E_\nu$ is the energy of the incoming neutrinos, $T_{nr}$ is the nuclear recoil energy,  $|\vec{q}|=\sqrt{2m_\mathcal{N} T_{nr}}$ is magnitude of the momentum transfer, and $m_\mathcal{N}$ is the mass of the target nucleus. The factor $Q^\text{SM}_V$ represents the weak charge of the nucleus, which can be defined as
\begin{eqnarray}
	Q^\text{SM}_V = g^p_VZ + g^n_VN,
\end{eqnarray}
where $g^p_V = (2g^u_V + g^d_V)$ and $g^n_V = (g^u_V + 2g^d_V)$. Here, $g^u_V$ and $g^d_V$ are the neutral current coupling constants for the ‘up’ and ‘down’ quarks, respectively. In terms of the weak mixing angle $\theta_W$ at low momentum transfer with $\sin^2\theta_W = 0.23857$ \cite{PDG2020}, we have
\begin{eqnarray}
	\begin{split}
		g^p_V &= -2\sin^2\theta_W + \frac{1}{2} \approx 0.0229, \quad g^n_V &= -\frac{1}{2}.
	\end{split}
\end{eqnarray}

We use the Klein-Nystrand parametrization \cite{Klein1999} for the form factor $F(|\vec{q}|^2)$. 
This form factor can be written as
\begin{eqnarray}
	F(|\vec{q}|^2) =3 \frac{J_1(|\vec{q}|R_A)}{|\vec{q}|R_A}\left(\frac{1}{1+|\vec{q}|^2a_k^2}\right),
\end{eqnarray}
where $J_1$ is the spherical Bessel function of order one, $R_A=1.23 A^{1/3}$ is the nuclear root mean square radius in fm for atomic mass number $A$, and $a_k=0.7$ fm. For the small momentum transfer, the impact of the nuclear recoil is negligible.
%

\subsection{Sterile neutrino dipole portal}
The presence of a transition magnetic moment between three active neutrinos and one sterile neutrino leads to the possibility that the sterile neutrino $\nu_4$ could be produced by neutrino beams electromagnetically up-scattered on nuclei.
Such interaction can be described by the effective Lagrangian \cite{Bolton2022}
\begin{eqnarray} \label{eq:lag}
	\mathcal{L}_\text{int} \supset \frac{\mu_{\nu_{\ell 4}}}{2} \bar{\nu}_{\ell L} \sigma^{\mu\nu} P_R \nu_4 F_{\mu\nu} + h.c.,
\end{eqnarray}
where $\mu_{\nu_{\ell 4}}$ is the active-sterile transition dipole coupling, $\nu_4$ is the sterile neutrino field, $\nu_{\ell L}$ is an SM left-handed (active) neutrino field of flavor $\ell=e, \mu, \tau$ and $F_{\mu\nu}$ is the usual electromagnetic field tensor. 
Here, the active and sterile neutrinos can either be Dirac or Majorana neutrinos. 
It should be noted that the Lagrangian \eqref{eq:lag} is only valid at energies below the electroweak (EW) scale. The CE$\nu$NS process occurs at energies well below the EW scale and thus it remains applicable.
%

\begin{figure}[h]
	\centering
	\includegraphics[scale=0.9]{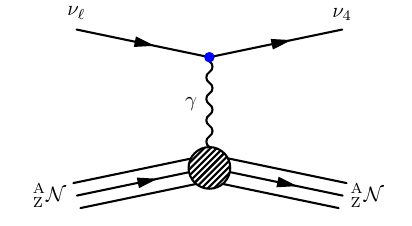}
	\caption{Representative diagram for the upscattering of $\nu_\ell \mathcal{N}$ → $\nu_4 \mathcal{N}$. The blue vertex denotes the neutrino dipole portal, which allows for a neutrino to up-scatter off a nucleus to a heavy neutral lepton state.}
	\label{fig:upscattering}
\end{figure}
The active-sterile transition dipole coupling $\mu_{\nu_{\ell 4}}$ leads to upscattering process $\nu_\ell \mathcal{N}$ → $\nu_4 \mathcal{N}$ \cite{Giunti:2015,Magill:2018jla} as shown in Fig. \ref{fig:upscattering}.
An incoming active neutrino $\nu_\ell$ exchanges a photon with a target nucleus $\mathcal{N}$ and up-scatters to a sterile neutrino $\nu_4$. The matrix element of this process can be written as \cite{Bolton2022}
\begin{eqnarray} \label{eq:amp}
	i\mathcal{M} = (\mu_{\nu_{\ell 4}})^* [\bar{u}_{\nu_4} \sigma^{\mu\nu} P_L q_\nu u_{\nu_\ell}] (\frac{-ig_{\mu\lambda}}{q^2}) j_\mathcal{N}^\lambda,
\end{eqnarray}
with the hadronic current of the nucleus
\begin{eqnarray} \label{eq:jamp}
	j_\mathcal{N}^\lambda = -ieZ (\bar{u}_\mathcal{N} \gamma^\lambda u_\mathcal{N}) F(|\vec{q}|^2),
\end{eqnarray}
where the target nucleus is considered as a spin-1/2 particle.
The $q$ denotes to the four-momentum of the photon. 
The differential cross section for neutrino scattering off a spin one-half nucleus can be calculated by taking the absolute square of the matrix element \eqref{eq:amp}, averaging over the possible initial spin states, and summing over the final spin states, to obtain
\begin{eqnarray}
	\begin{split}
		\biggl[\frac{d\sigma (E_\nu, T_{nr})}{dT_{nr}} &\biggr]^{\nu_\ell \mathcal{N} \rightarrow \nu_{4} \mathcal{N}}_ {\text{spin}-1/2}= \frac{\pi \alpha_{\mathrm{EM}}^2}{m_e^2} \left| \frac{\mu_{\nu_{\ell 4}}}{\mu_B} \right|^2 Z^2 F^2(|\vec{q}|^2)  \Bigg[\frac{1}{T_{nr}} \\ & - \frac{1}{E_\nu} -  \frac{m_4^2}{2T_{nr}E_\nu m_\mathcal{N} } \left(1-\frac{T_{nr}}{2E_\nu}+\frac{m_\mathcal{N}}{2E_\nu}\right) \\
		&- \frac{m_4^4}{8m_\mathcal{N} T_{nr}^2 E_\nu^2} \left(1-\frac{T_{nr}}{m_\mathcal{N}} \right) \Bigg],
		\end{split}
\end{eqnarray}
where subdominant scattering via the nuclear magnetic dipole moment is neglected. The factor of $\pi \alpha_{\mathrm{EM}}/m_e^2$ arises from a usual normalization of magnetic moments according to the Bohr magneton $\mu_B$. This result agrees with those found in Refs.~\cite{Bolton2022,Li2022}. For $m_4=0$, the above expression becomes the conventional active neutrino magnetic moment cross-section (see, Ref.~\cite{Vogel:1989iv}). 
It should be noted that the sterile neutrino mass must satisfy the following kinematic constraint
\begin{eqnarray}\label{eq:m4const}
	m_4^2 \leq 2m_\mathcal{N} T_{nr} \left(\sqrt{\frac{2}{m_\mathcal{N} T_{nr}}}E_\nu -1 \right).
\end{eqnarray}
On the other hand, assuming a spin-0 nuclei, the differential cross section reads 
\begin{eqnarray} 
	\begin{split}
		\biggl[\frac{d\sigma }{dT_{nr}} \biggr]_{\text{spin-0}}
		&=\biggl[\frac{d\sigma }{dT_{nr}} \biggr]_ {\text{spin-1/2}}+ \frac{\pi \alpha_{\mathrm{EM}}^2}{m_e^2} \left| \frac{\mu_{\nu_{\ell 4}}}{\mu_B} \right|^2 Z^2 F^2(|\vec{q}|^2) \\ & \times \Bigg[\frac{T_{nr}}{4 E_\nu^2} -\frac{m_4^2}{8m_\mathcal{N}E_\nu^2} \left(1+\frac{m_4^2 }{m_\mathcal{N} T_{nr}} \right) \Bigg],
	\end{split}
\end{eqnarray}
where the last two terms showing the difference from those of spin-1/2 nucleus are proportional to $\frac{1}{E_\nu^2}$ and  $\frac{1}{m_\mathcal{N} E_\nu^2}$, respectively. The contribution of a transition magnetic moment between active and sterile neutrinos is incoherently added with the SM case.

\section{Data Analysis Details}
\label{section:data}
\subsection{Differential Rate}
\begin{figure}[th]
	\centering	\includegraphics[scale=0.43]{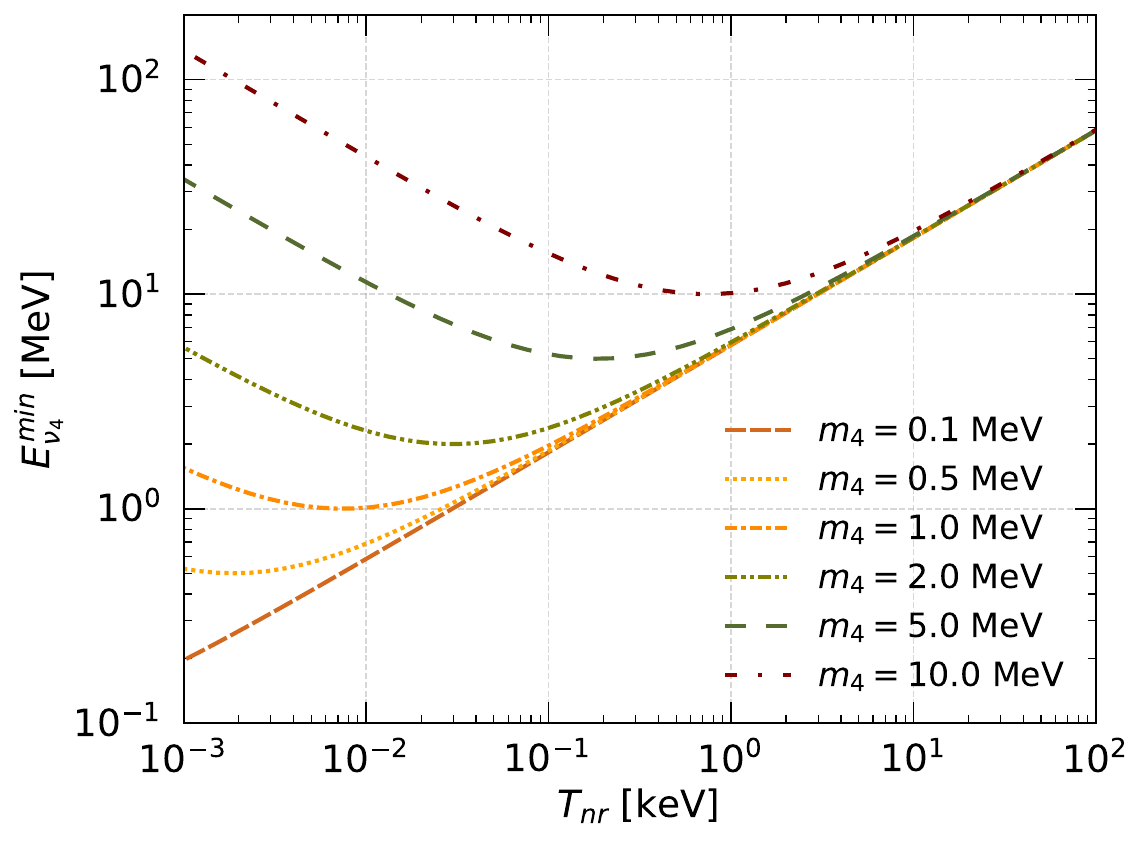}
	\caption{The minimum neutrino energy  $E_{\nu_4}^{min}$ as a function of the nuclear recoil energy for several sterile neutrino masses. }
	\label{fig:Emin}
\end{figure}

In the present work, we focus on the CE$\nu$NS process induced by solar neutrinos in the presence of a transition magnetic moment between active and sterile neutrinos. 
The corresponding nuclear recoil event is given by
\begin{align}
	\frac{dR}{dT_{nr}} = \frac{\epsilon}{m_T}
	\int_{E_{\nu}^\text{min}}^{E_{\nu}^\text{max}} dE_\nu \frac{d  \Phi_{\nu_{\ell}}^{i} (E_\nu)}{dE_{\nu}} \biggl[\frac{d\sigma(E_\nu, T_{nr})}{dT_{nr}}\biggr]  ,
\end{align}
where $\epsilon$ is experimental exposure, $m_T$ is the target mass and $\frac{d  \Phi_{\nu_{\ell}}^{i} (E_\nu)}{dE_{\nu}}$ is the solar neutrino flux per cm$^2$ per second. 
The integration is taken from the minimum neutrino energy $E_{\nu}^\text{min}$ to the maximum neutrino energy $E_{\nu}^\text{max}$. The minimum neutrino energy is given by
\begin{align}
	E_{\nu}^\text{min} = \frac{T_{nr}}{2}\left(1+\sqrt{1+\frac{2m_\mathcal{N}}{T_{nr}}} \right)
\end{align}
for the active neutrino, and
\begin{eqnarray}
	E_{\nu_4}^\text{min}=\frac{m_4^2 + 2m_\mathcal{N} T_{nr}}{2(\sqrt[]{T_{nr}(T_{nr}+2m_\mathcal{N}}) - T_{nr})}
	\label{eq:emin_sterile}
\end{eqnarray}
for the sterile neutrino, which is higher than the active neutrino case.  Neutrino sources with higher neutrino energy can produce sterile neutrinos with larger $m_4$. As can be seen from Fig. \ref{fig:Emin}, the $E^\text{min}_{\nu_4}$ continuously decreases with the increment of the recoil energy until it reaches the extreme value $E^\text{min}_{\nu_4} = m_4 + m_4^2/2m_\mathcal{N}$.

\begin{figure}[htb]
	\centering	\includegraphics[scale=0.43]{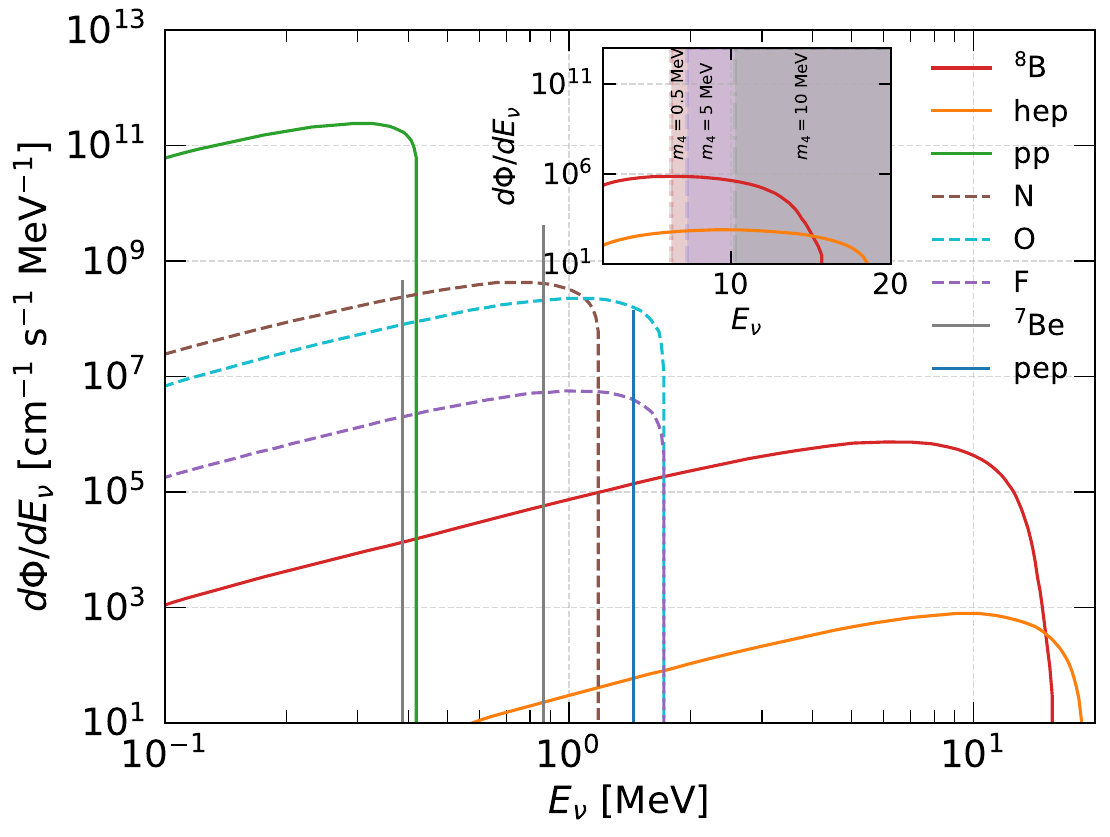}
	\caption{Solar neutrino flux components for the pp chain and CNO cycles derived from the high-metallicity solar neutrino model BS05(OP).}
	\label{fig:solarflux}
\end{figure}

The solar neutrino fluxes $\Phi_{\nu_{\ell}}^{i}$  are given by the standard solar model (SSM) BS05(OP) \cite{Bahcall:2004mq,Bahcall:2004pz}. Figure \ref{fig:solarflux} shows solar neutrino flux components coming from different reactions taking place at the core of the Sun. Solar neutrinos span a wide energy range of up to almost 15 MeV. 
In our analysis, we consider the $^{8}$B and $hep$ solar neutrino spectra, which provide the main contributions to the event rates of the experiment in question. These arise
from two nuclear processes $^{8}\text{B} \to ^{8}\text{Be}^*+e^+ +\nu_e$ and $^{3}\text{He} +p \to ^{4}\text{He}^*+e^+ +\nu_e$ in the Sun.
In the inset figure, we also show the region of minimum neutrino energies on the considered solar fluxes for sterile neutrino masses of 0.5 MeV, 5 MeV, and 10 MeV. These correspond to $E_\nu=$ 6.28 MeV, 7.27 MeV, and 10.26 MeV, respectively. The minimum neutrino energy for the sterile case approaches those of the active neutrino as $m_4$ becomes smaller.

The solar neutrinos oscillate as they propagate from the Sun to the Earth. They arrive at a detector on Earth as a mixture of $\nu_{e}$, $\nu_{\mu}$, and $\nu_{\tau}$. We take the survival probabilities for each flavor to be 
\begin{eqnarray}
	\Phi_{\nu_{e}}^{i}&=&\Phi_{\nu_{e}}^{i\,\odot} P_{ee},  \\
	\Phi_{\nu_{\mu}}^{i}& =& \Phi_{\nu_{e}}^{i\,\odot} \left( 1 - P_{ee} \right) \cos^2\vartheta_{23},\\
	\Phi_{\nu_{\tau}}^{i} &= &\Phi_{\nu_{e}}^{i\,\odot} \left( 1 - P_{ee} \right) \sin^2\vartheta_{23},
	\label{eq:solflux}
\end{eqnarray}
where $\Phi_{\nu_{e}}^{i\,\odot}$ is the electron-neutrino flux produced by thermonuclear reactions in the core of the Sun, with $i = hep$ and $ ^{8}\text{B}$, etc. The $P_{ee}$ is the survival probability of $\nu_{e}$ on the Earth, which can be	given by \cite{Maltoni:2015kca}
\begin{eqnarray}
	\begin{split}
		P_{ee} = & (c_{13}^2 {c_{13}^m}^2)\left( \frac{1}{2} - \frac{1}{2} \cos 2\vartheta_{12}^m \cos 2\vartheta_{12} \right) \\ &+ (s_{13}^2 {s_{13}^m}^2)
		\label{Pee}
	\end{split}
\end{eqnarray}
where $c_{13}=\cos \vartheta_{13}$, $s_{13}=\sin \vartheta_{13}$ and the label $m$ represents the matter effect. In this equation $\cos 2\vartheta_{12}^m$ is the matter angle. 
We consider the day-night asymmetry due to the Earth matter effect in the calculation of the survival probabilities. We take the normal-ordering neutrino oscillation parameters from the latest 3-$\nu$ oscillation of NuFit-5.3, without the Super-Kamiokande atmospheric data \cite{Esteban:2020cvm}. These are $\sin^2\vartheta_{12} = 0.307^{+0.012}_{-0.011}$, $\sin^2\vartheta_{23} = 0.572^{+0.018}_{-0.023}$, and $\sin^2\vartheta_{13} = 0.02203^{+0.00056}_{-0.00058}$, as well as $\Delta m_{12}^2 = 7.41^{+0.21}_{-0.20} \times 10^{-5} \text{ eV}^2$ and $\Delta m_{3\ell}^2 = 2.511^{+0.027}_{-0.027} \times 10^{-3} \text{ eV}^2$.

It should be noted that the data are measured in units of electron equivalent energy $T_{ee}$. To convert the nuclear recoil energy $T_{nr}$ into $T_{ee}$, a quenching factor $Y(T_{nr})$ is used, which relates these two quantities as follows:
\begin{align}
	T_{ee} = Y(T_{nr}) T_{nr}.
\end{align}
For this aim, we use the Linhard quenching factor \cite{Lindhard:1963}
\begin{align}
	Y(T_{nr}) = \frac{kg(\epsilon)}{1+kg(\epsilon)} 
	\label{eq:Linhard_quench}
\end{align}
where $g(\epsilon)= 3\epsilon^{0.15} + 0.7 \epsilon^{0.6} + \epsilon$ and $\epsilon = 11.5 {Z}^{-7/3} T_{nr}$. We consider $k=0.162$ to closely match the recent measurement in the low-energy range \cite{Bonhomme:2022lcz}. This quenching factor is acceptable for $T_{nr}>0.254\text{ keV}$. 
For lower recoil-energy, we consider another quenching factor
\begin{align}
	Y(T_{nr}) = 0.18\left[1-e^{\left(\frac{15-T_{nr}}{71.03}\right)}\right],
	\label{eq:lowquench}
\end{align}
obtained from the ``high'' ionization-efficiency model for Ge target, which is acceptable in the range of $0.015 \text{ keV}<T_{nr}<0.254 \text{ keV}$ \cite{Essig:2018tss}.
Consequently, in terms of the electron equivalent energy, the differential rate can be written as
\begin{align}
	\frac{dR}{dT_{ee}} = \frac{dR}{dT_{nr}} \frac{1}{Y(T_{nr})+T_{nr} \frac{dY(T_{nr})}{dT_{nr}}}.
\end{align}
The effect of the quenching factor on the conversion between the nuclear recoil energy and its electron equivalency is shown in Fig. 2(b) of Ref. \cite{Demirci:2024}. In this conversion process, approximately $80\%$ of the initial energy is lost in dissipative processes and, thus, not accessible. 
We also note that there exist other measurements in the choice of quenching factors \cite{Collar:2021fcl}.

\subsection{$\chi^2$ Function}
In this work, we investigate the sensitivity to probe the transition neutrino magnetic moment on the CE$\nu$NS process induced by solar neutrinos.
In order to do that, we use the current data from CDEX-10 experiment \cite{CDEX:2022mlp} and the pull approach of the $\chi^2$ function given as \cite{Fogli:2002pt}
\begin{eqnarray}
	\begin{split}
		\chi^2 = \mathrm{min}_{(\xi_j)} \sum_{i=1}^{20} &\Bigg( \frac{ R_\text{obs}^{i} - R_\text{exp}^{i} - B - \sum_j \xi_j c_{j}^i }{\Delta^{i}}\Bigg)^2  \\
		&+  \sum_j \xi_j^2.
	\end{split}
\end{eqnarray} 
The function is minimized to pull parameters $\xi_j$ for the $j$-th neutrino flux. The terms $R_\text{obs}^i$ and $R_\text{exp}^i$ represent the $i$-th energy bin of the observed and expected (SM plus new physics contribution) event rates, respectively. The effect of efficiency has been considered in $R_\text{exp}^i$. This factor comes from the combination of the trigger efficiency and the physics-noise (PN) cut efficiency of the experiment \cite{CDEX:2018lau}. The term $\Delta^i$ represents the experimental uncertainty for the $i$-th energy bin, which includes statistical and systematic uncertainties \cite{CDEX:2018lau}.
Finally, the factor $c_j^i$ denotes the solar neutrino flux uncertainty.

\subsection{Projected Sensitivities}
Upcoming DD experiments are entering the multiton phase. Many such experiments have the potential for detecting CE$\nu$NS and exploring new physics scenarios. Experiments with xenon targets have already been able to observe the CE$\nu$NS process with solar neutrinos, such as PandaX-4T \cite{PandaX:2018wtu,PandaX:2024muv}, and XENONnT \cite{XENON:2020kmp, XENON:2024}, 
while a future DARWIN \cite{DARWIN:2016hyl} facility is planned to deploy 50 tons of liquid xenon target. Furthermore, next-generation low-scale solid material targets using silicon or germanium are anticipated to reach low threshold detectors. Some experiments with this purpose are EDELWEISS \cite{EDELWEISS:2022ktt}, Super CDMS \cite{SuperCDMS:2016wui}, and SENSEI \cite{SENSEI:2020dpa}. These are expected to improve constraints on low-mass WIMPs and to detect extremely low recoil energy events of solar neutrino CE$\nu$NS that may allow more severe tests for BSM physics.

The CDEX facility is currently developing its third phase, the so-called CDEX-50 \cite{Geng:2023yei}, as a next generation project. In this phase of the experiment, it is planned to have a detector array consisting of $50$ kg of high-purity germanium. It is expected to reduce the background to about $0.01 \text{ events }$ $ \text{keV}^{-1}\text{kg}^{-1} \text{day}^{-1}$ with the objective exposure goal of $150 \text{ kg}\cdot\text{year}$ and $160 \text{eVee}$ analysis threshold \cite{Geng:2023yei}. Moreover, the ultimate goal of the experiment is to set up a ton-scale mass Ge detector based on the PCGe detector \cite{CDEX:2013kpt}. It is expected to run in the near future, reaching a low sub-keV energy threshold. The upgrade is anticipated to improve the detectability of CE$\nu$NS by solar neutrinos with more precise data. 

In this regard, we construct three scenarios to offer perspective for future studies. We called these ``next-generation" and ``future" scenarios, where the latter case consists of two further scenarios which we call ``future 1" and ``future 2". These scenarios are based on the projected development of the DD facilities mentioned above. The ``next-generation" scenario is configured to have exposure of 150 $\text{kg}\cdot \text{ year}$, while both ``future" cases have exposure of 1.5 $\text{ton}\cdot \text{ year}$. In the ``next-generation" and ``future 1" scenarios, the nuclear recoil energy threshold is set to be 1 keVnr, while it is set to  0.1 keVnr in the ``future 2" scenario. The target mass for all the scenarios is considered to be $50$ kg. In each case, a flat background of $0.01 \text{ events } \text{keV}^{-1}\text{kg}^{-1} \text{day}^{-1}$ is assumed. These configurations are supposed to reduce the uncertainty by a factor of 10 \% and 1 \% for the next-generation and future scenarios, respectively. With the help of these projections, we intend to investigate the relationship between the sterile neutrino mass bounds and experimental accuracy that may provide intuitive scaling for DD advancements.
\section{Results}\label{sec:results}
In this section, we present our numerical results for the differential cross section, the expected event rates and the data analysis. We provide differential rate predictions of the CE$\nu$NS process as indicative of the emergence of the active-sterile neutrino transition magnetic with the solar neutrinos. 
For the solar neutrino flux, we consider the sum of the $^8\mathrm{B}$ and ${hep}$ components, unless otherwise stated. We use the notation $\mu_{\nu_{\ell 4}}$ to represent the flavor-independent effective transition magnetic moment where all active neutrinos are also considered. Flavor-dependent cases are evaluated by taking solar neutrino survival probabilities into consideration in which the index $\ell$ could be $e,\mu$ or $\tau$. 

\subsection{Differential cross section}
\begin{figure}[h]
	\centering
	\includegraphics[scale=0.4]{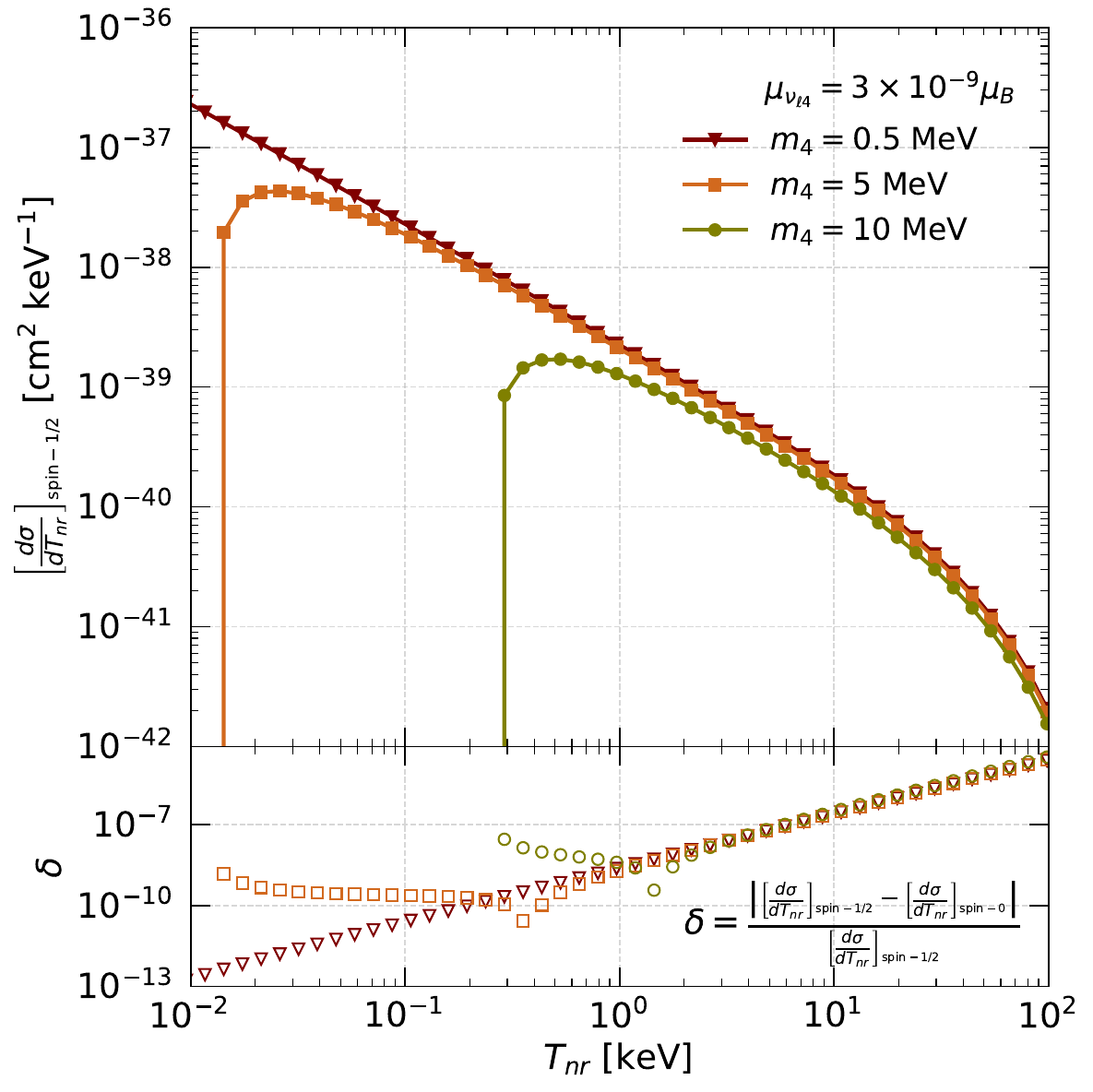}
	\caption{The effect of assuming it as a spin-0 or spin-1/2 nucleus on the differential cross-section for the upscattering of $\nu_\ell \mathcal{N}$ → $\nu_4 \mathcal{N}$.}
	\label{fig:delta_xsec_lin}
\end{figure}

To show the effect of assuming nucleus as a spin-0 or spin-1/2 particle, we plot in Fig. \ref{fig:delta_xsec_lin} the differential cross section with a spin-1/2 nucleus and the relative difference $\delta$ between spin-0 and spin-1/2 cases defined by
\begin{eqnarray}
	\delta =\frac{\biggl|\biggl[\frac{d\sigma }{dT_{nr}} \biggr]_ {\text{spin-1/2}}-\biggl[\frac{d\sigma }{dT_{nr}} \biggr]_ {\text{spin-0}}\biggr|}{\biggl[\frac{d\sigma }{dT_{nr}} \biggr]_ {\text{spin-1/2}}}.
\end{eqnarray}
\begin{figure*}[ht]
	\centering
	\includegraphics[scale=0.43]{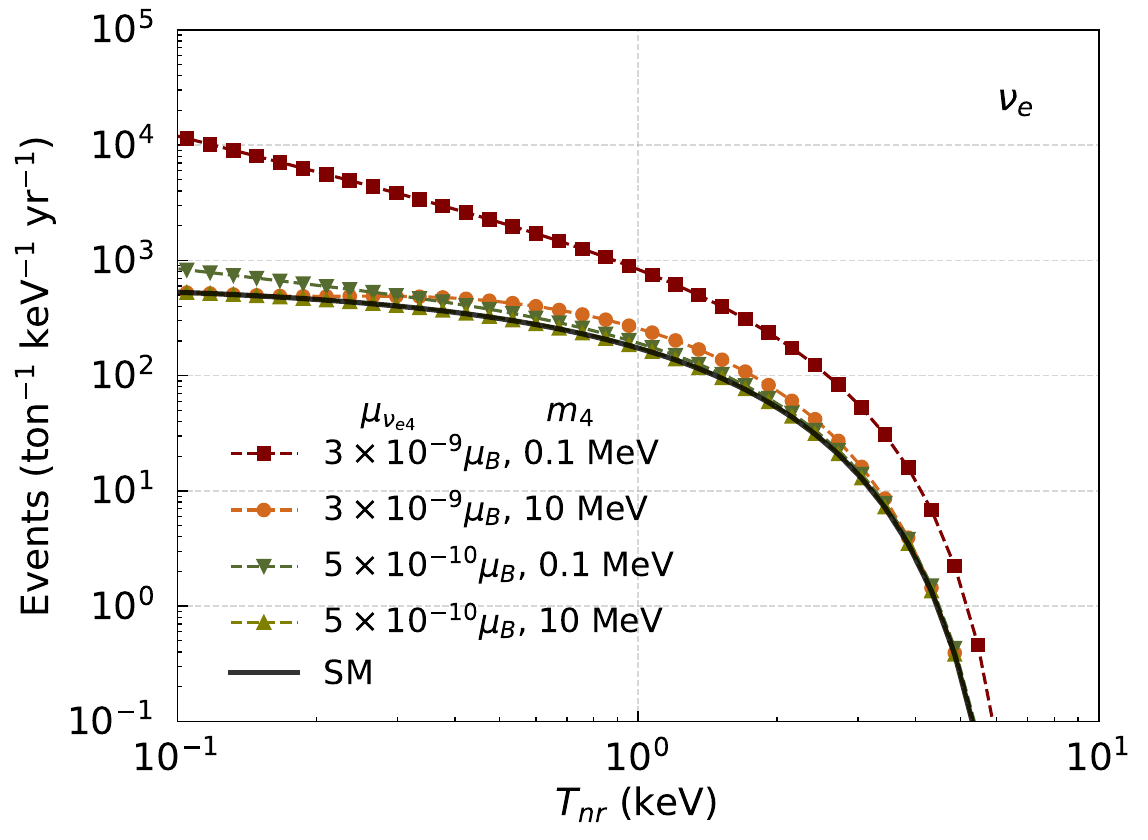}
	\includegraphics[scale=0.43]{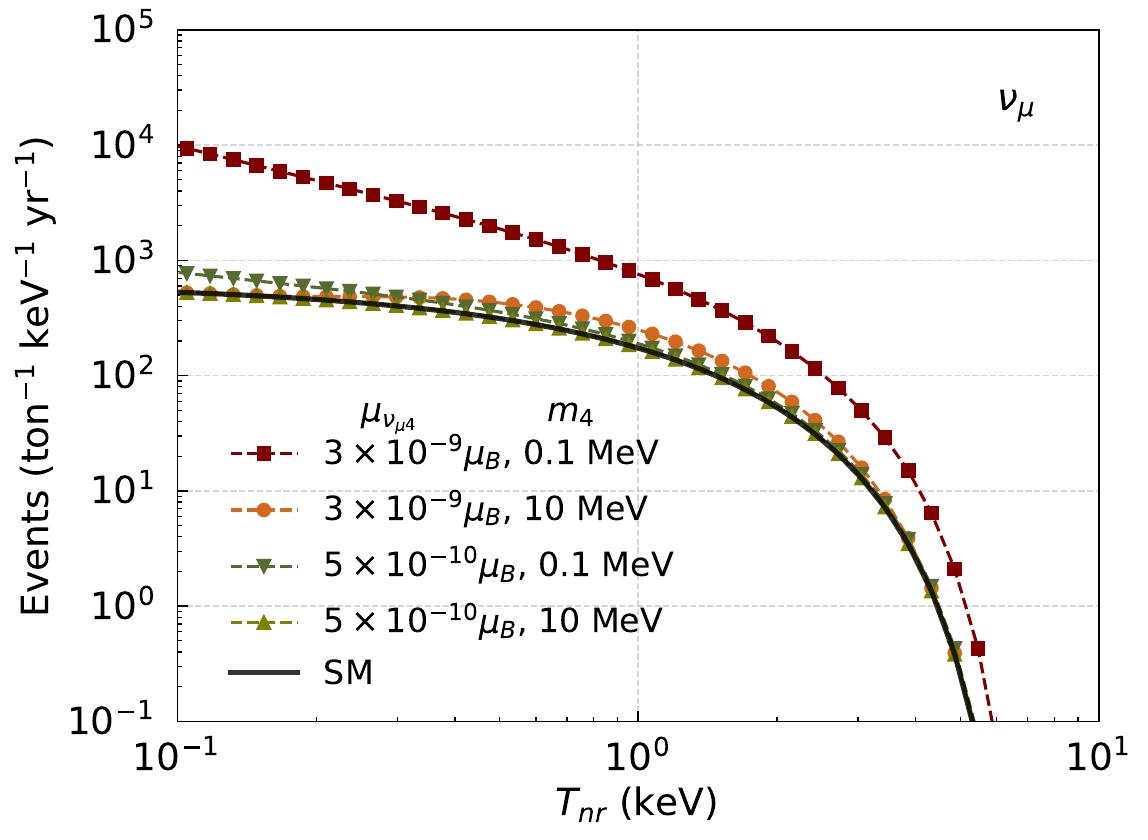}
	\\
	\hspace{10.4mm}	(a) \hspace{80.8mm} (b)
	\includegraphics[scale=0.43]{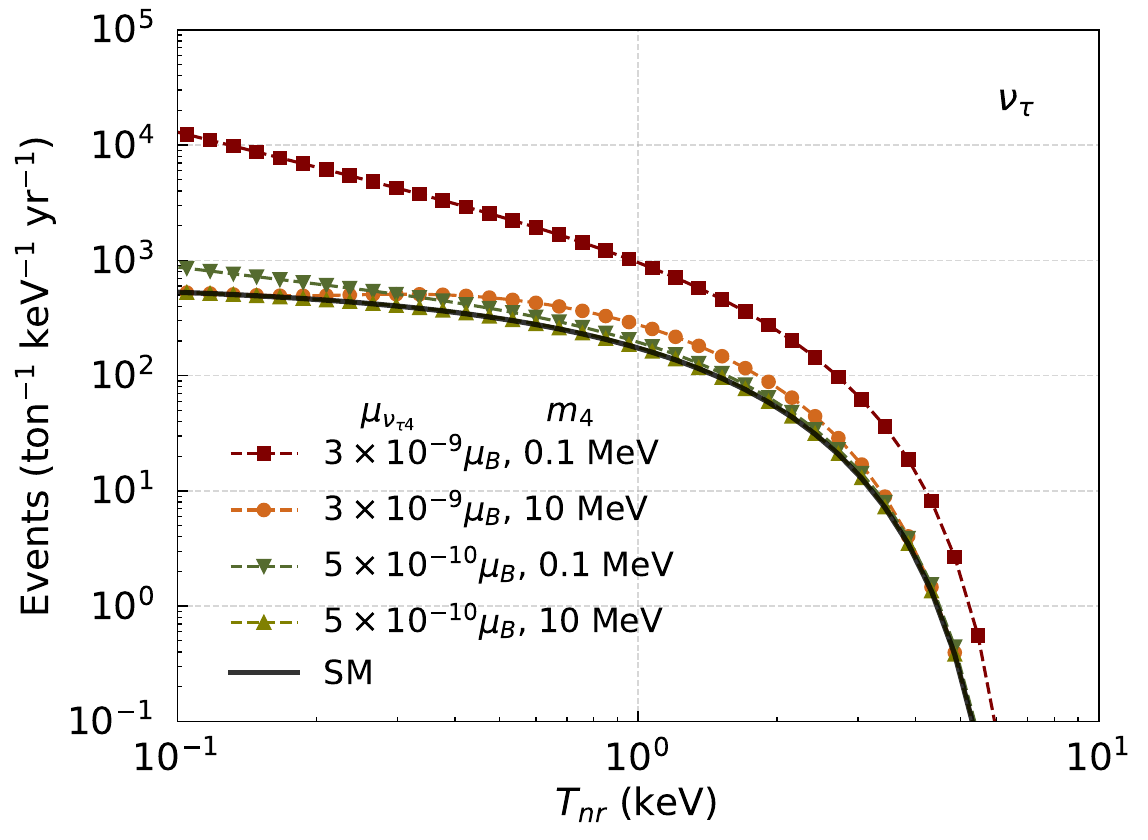}
	\includegraphics[scale=0.43]{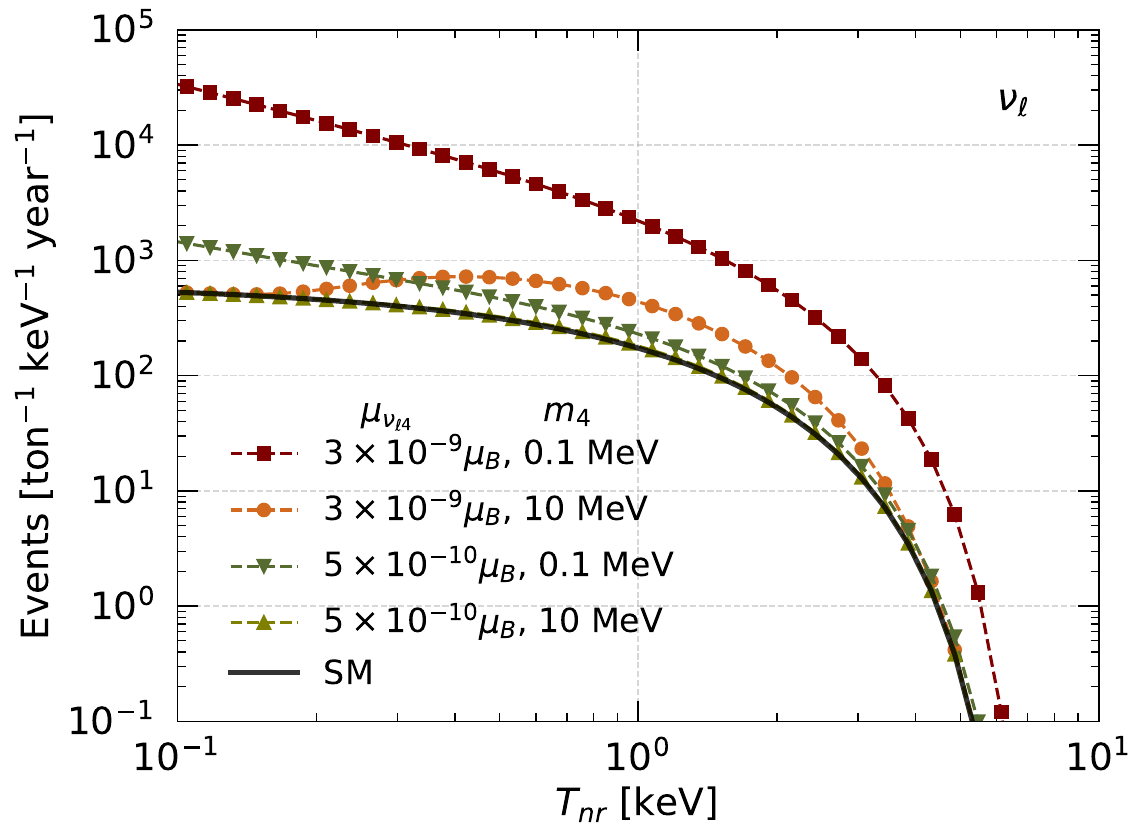}
	\\
	\hspace{10.4mm}	(c) \hspace{80.8mm} (d)
	\caption{Predicted event rates as a function of the nuclear recoil energy induced by solar neutrinos in the presence of the active (a) $\nu_e$, (b) $\nu_\mu$, (c) $\nu_\tau$, and (d) flavor-independent effective $\nu_{\ell}$-sterile neutrino transition magnetic moments.  The event rates are shown as the SM plus nonstandard contributions from transition magnetic moment.
		} 
	\label{fig:eventrate}
\end{figure*}

We set the benchmark points as $m_4=0.5~$MeV, $5$ MeV, and $10$ MeV for a fixed of $\mu_{\nu_{\ell 4}}=3 \times 10^{-9} \mu_B$ and $E_\nu = 10$ MeV. It is clear that the differential cross sections have the same order of magnitude for both spin-0 and spin-1/2 nuclei.  
The cross sections increase quickly with the opening of the phase space for a given initial value of $T_{nr}$ and then decrease with increments of $T_{nr}$. This initial value varies depending on the value of $m_4$. 
As $T_{nr}$ increases, all curves begin to overlap, especially for $T_{nr}>10$ keV.  However, the relative difference $\delta$ is in order of $10^{-5}$ to $10^{-13}$ for the considered interval of $T_{nr}$, as shown in the bottom panel of Fig. \ref{fig:delta_xsec_lin}. This implies that the different spin states do not significantly alter the result. We note that a dip occurs in $\delta$ at the threshold values of $T_{nr}$ where the cross section of the spin-0 nucleus becomes larger than those of spin-1/2. For $m_4=0.5~$MeV, $5$ MeV, and $10$ MeV, these threshold values are around $T_{nr}=0.0038~$keV, $0.39$ keV, and $1.6$ keV, respectively.

\subsection{Expected event spectra}
We present the predicted event rates as a function of the nuclear recoil energy induced by solar neutrinos for both SM and the presence of active-sterile neutrino transition magnetic moments in Fig.~\ref{fig:eventrate}. 
%
\begin{figure*}[ht]
	\centering
	\includegraphics[scale=0.43]{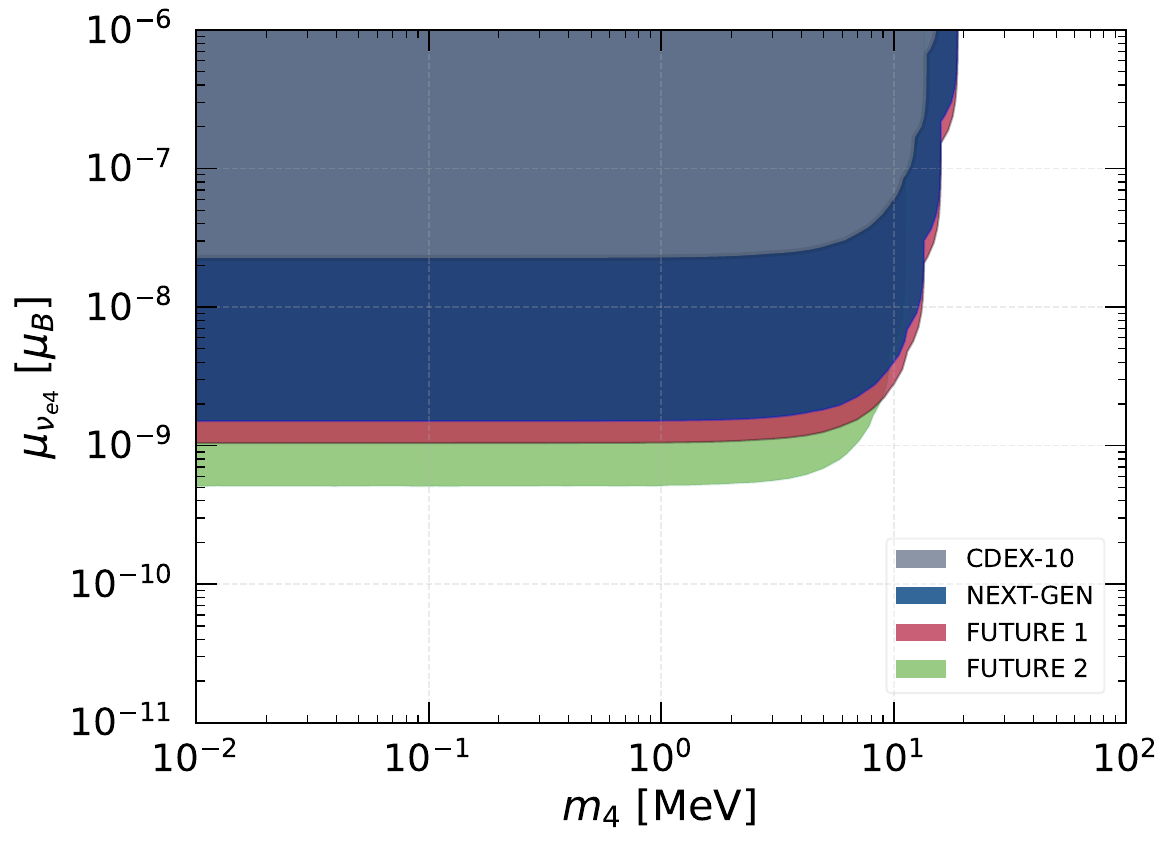}
	\includegraphics[scale=0.43]{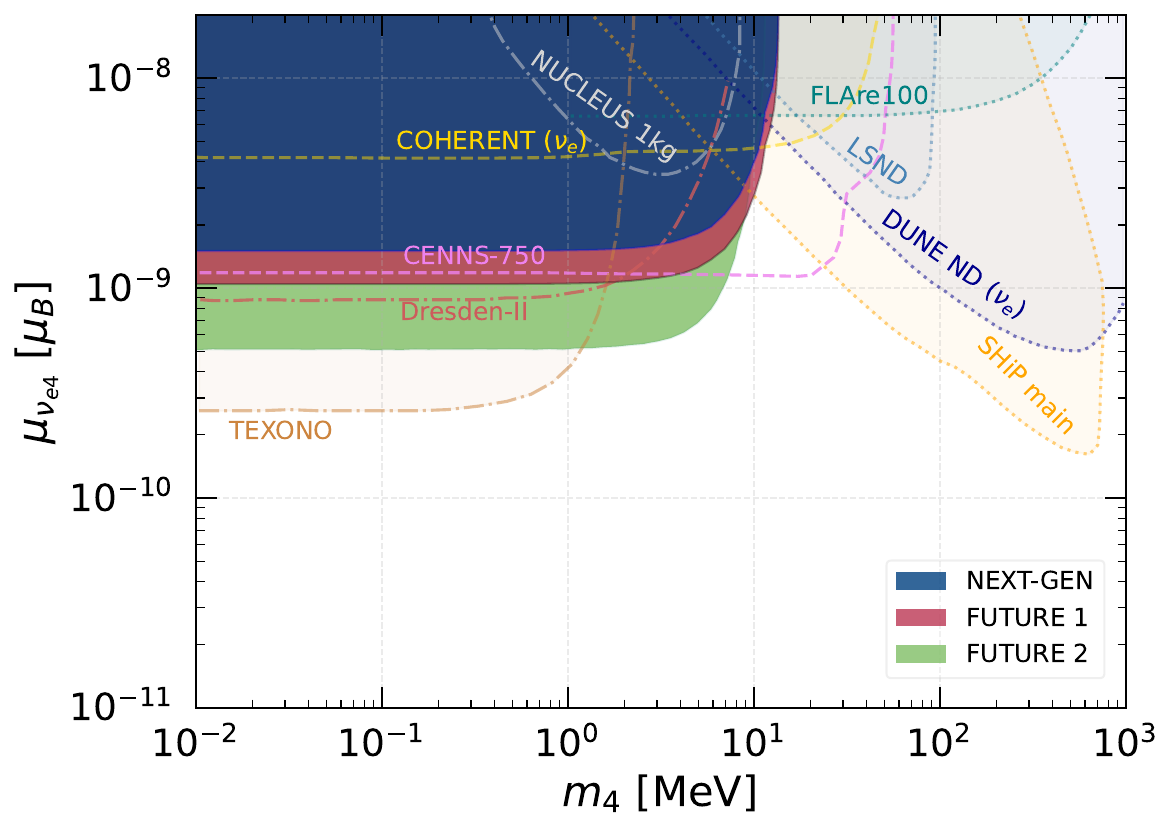}
	\\
	\hspace{10.4mm}	(a) \hspace{80.8mm} (b)
	\caption{(a) 90$\%$ C.L. (2 d.o.f.) exclusion regions on the plane of the active $\nu_e$-sterile neutrino transition magnetic moment vs sterile neutrino mass from the current CDEX-10 data, next-generation and future scenarios, and (b) comparison with other available experimental constraints (see the text for details).}
	\label{fig:B8hep_flav_e}
\end{figure*}
These are given in the unit of per ton per keV per year.  We show total number of event rates i.e., the SM plus non-standard (contribution of transition magnetic moment). 
We consider four benchmark points, which are set as two different transition magnetic moments of $3\times 10^{-9} \mu_B$ and $5\times 10^{-10} \mu_B$ together with a light sterile neutrino with a mass of $0.1 \text{ MeV}$ and a heavy one with $10 \text{ MeV}$.
Since the cross section of the process includes inversely dependent terms on the nuclear recoil energy, we can see that the event rates are enhanced in low nuclear recoil energy. This demonstrates the importance of observing low nuclear recoil thresholds in experiments. 
Also note that the flavor-independent case provides larger event predictions than those of flavor-dependent for all the chosen values due to the oscillation probability.
The light sterile mass significantly increases the predicted event rates at low recoil energies, which will be the main contribution to improvements in the search for transition magnetic moments.
One may notice that the distribution for $m_4 = 10$ MeV with $\mu_{\nu_\ell}=3\times 10^{-9} \mu_B$ slowly vanish as $T_{nr} \lesssim 0.2$ keV. 
This indicates that the heavier mass suppresses the signals and provides relatively insignificant effects to the standard prediction.

\subsection{Constraints on the sterile neutrino dipole portal}
We derive new limits on the dependence of the active-sterile neutrino transition magnetic moment on sterile neutrino mass from the CDEX-10 data for flavor dependent and flavor independent cases. 
We also provide projected sensitivities for next-generation and future scenarios by taking into account experimental developments discussed earlier. 
\begin{figure*}[ht]
	\centering
	\includegraphics[scale=0.43]{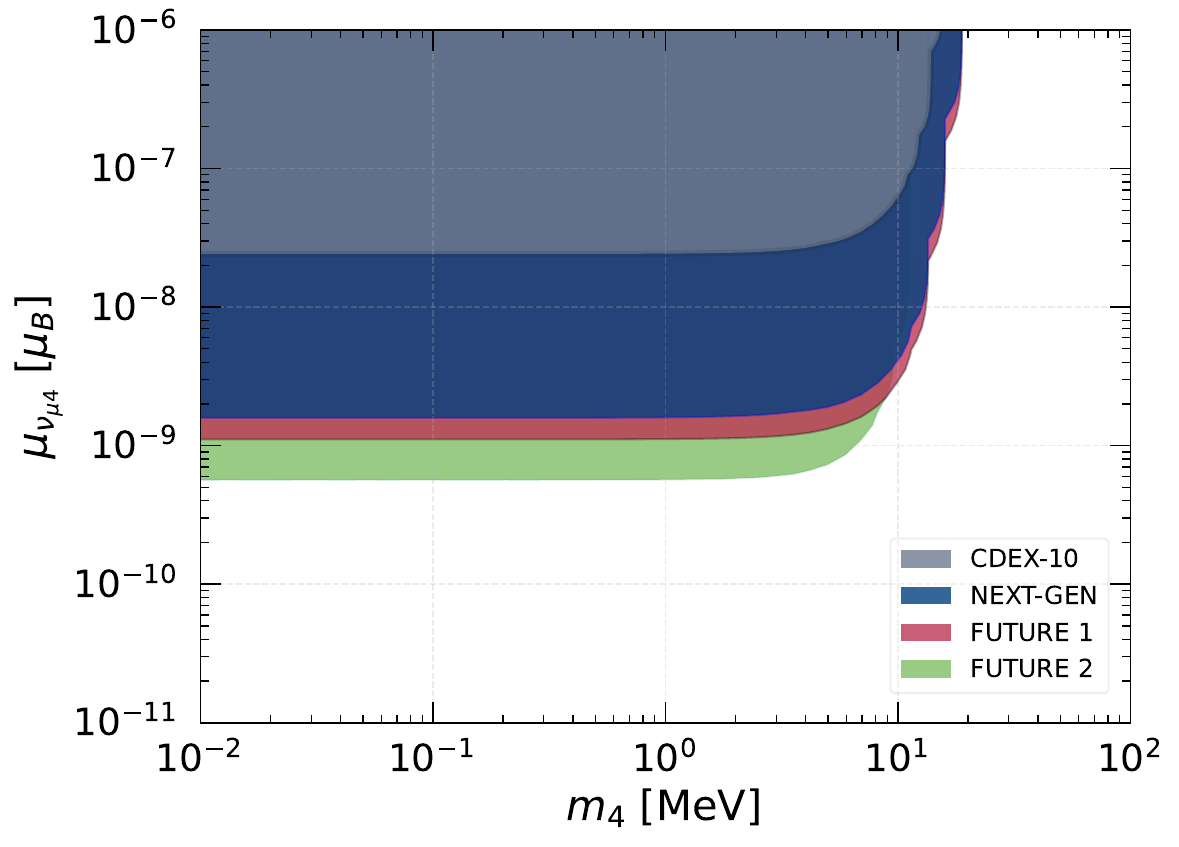}
	\includegraphics[scale=0.43]{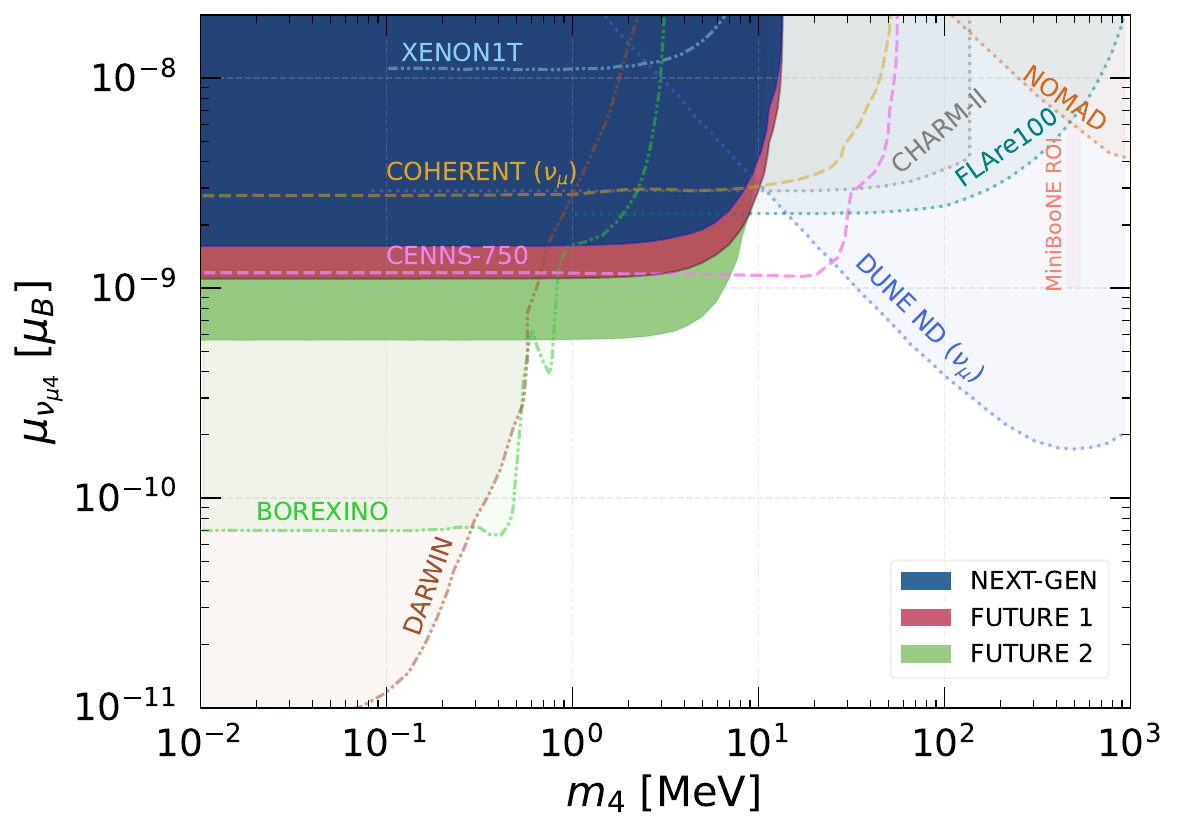}
	\\
	\hspace{10.4mm}	(a) \hspace{80.8mm} (b)
	\caption{(a) 90$\%$ C.L. (2 d.o.f.) exclusion regions on the plane of the active $\nu_\mu$-sterile neutrino transition magnetic moment vs sterile neutrino mass from the current CDEX-10 data, next-generation and future projected scenarios, and (b) comparison with other available experimental constraints (see the text for details).}
	\label{fig:B8hep_flav_mu}
\end{figure*}
We further compare our results with the available limits derived from SNS (COHERENT CsI+LAr in Ref.~\cite{DeRomeri:2022twg}), accelerator neutrino (LSND in Ref.~\cite{Magill:2018jla}), beam dump (CHARM in Ref.~\cite{Coloma:2017ppo}), nuclear reactor (TEXONO in Ref.~\cite{Miranda2021} and DRESDEN-II in Ref.~\cite{AtzoriCorona:2022} ), solar neutrino (BOREXINO in Ref.~\cite{Brdar:2020quo}), DD experiments (XENON1T in Refs.~\cite{Miranda2021,Brdar:2020quo}), and expected limits from future neutrino observatories (CENNS-750 in Ref.~\cite{Miranda2021}, DARWIN in Ref.~\cite{Brdar:2020quo}, NUCLEUS in Ref.~\cite{Bolton2022}, DUNE Near Detector (ND) \cite{Schwetz2021,Atkinson:2021rnp}, and SHiP \cite{Magill:2018jla}). 
For the specific flavor-dependent cases, we additionally consider the NOMAD limit derived in Ref.~\cite{Gninenko:1998nn}, MiniBooNe in Ref.~\cite{Magill:2018jla}, ALEPH, DONUT, and FLArE-100 limit in Ref.~\cite{Ismail:2021dyp}, as well as IceCube/DeepCore (DC) limits in Ref.~\cite{Coloma:2017ppo}. We include sensitivities from the cosmological study of Big Bang Nucleosynthesis (BBN) and Supernova 1987 A (SN1987A) ~\cite{Brdar:2020quo}. Furthermore, we show the limit from sterile neutrino decay $\nu_4 \rightarrow \nu \gamma$ \cite{Plestid:2020vqf}.
\begin{figure*}[ht]
	\centering
	\includegraphics[scale=0.43]{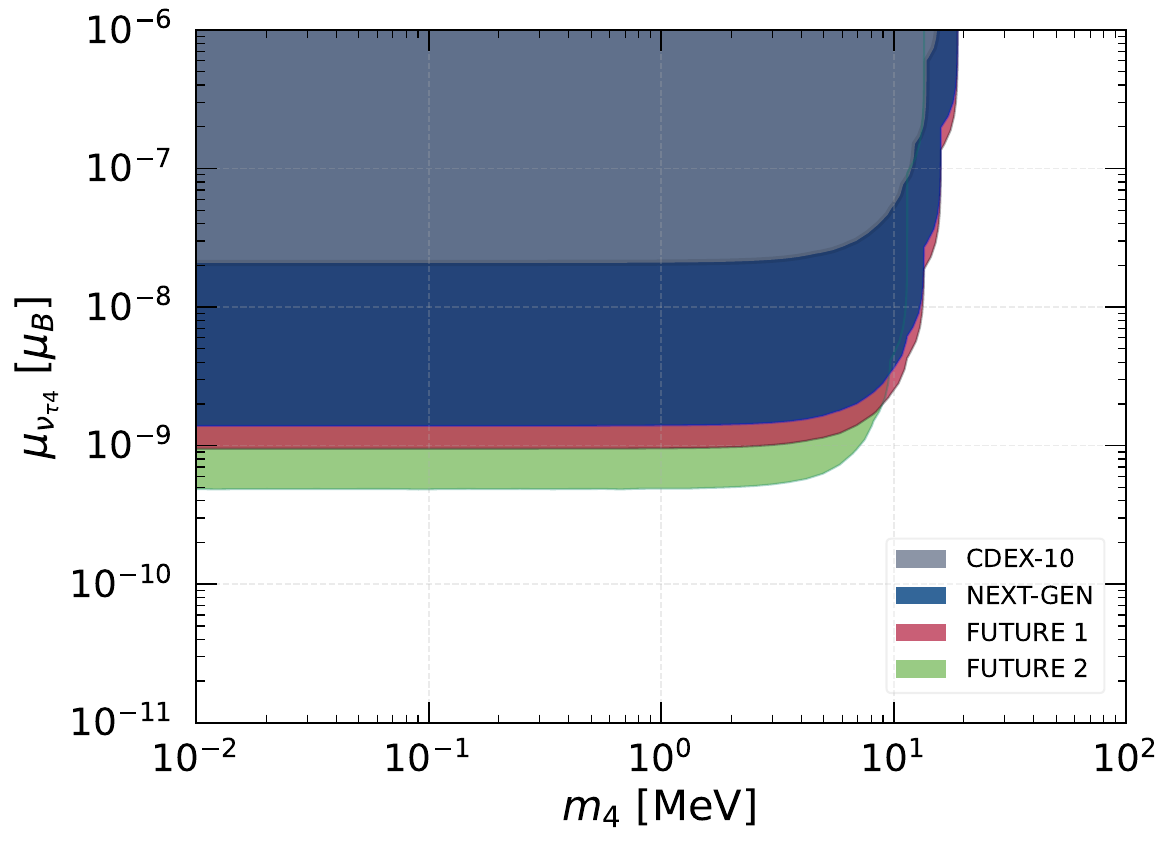}
	\includegraphics[scale=0.43]{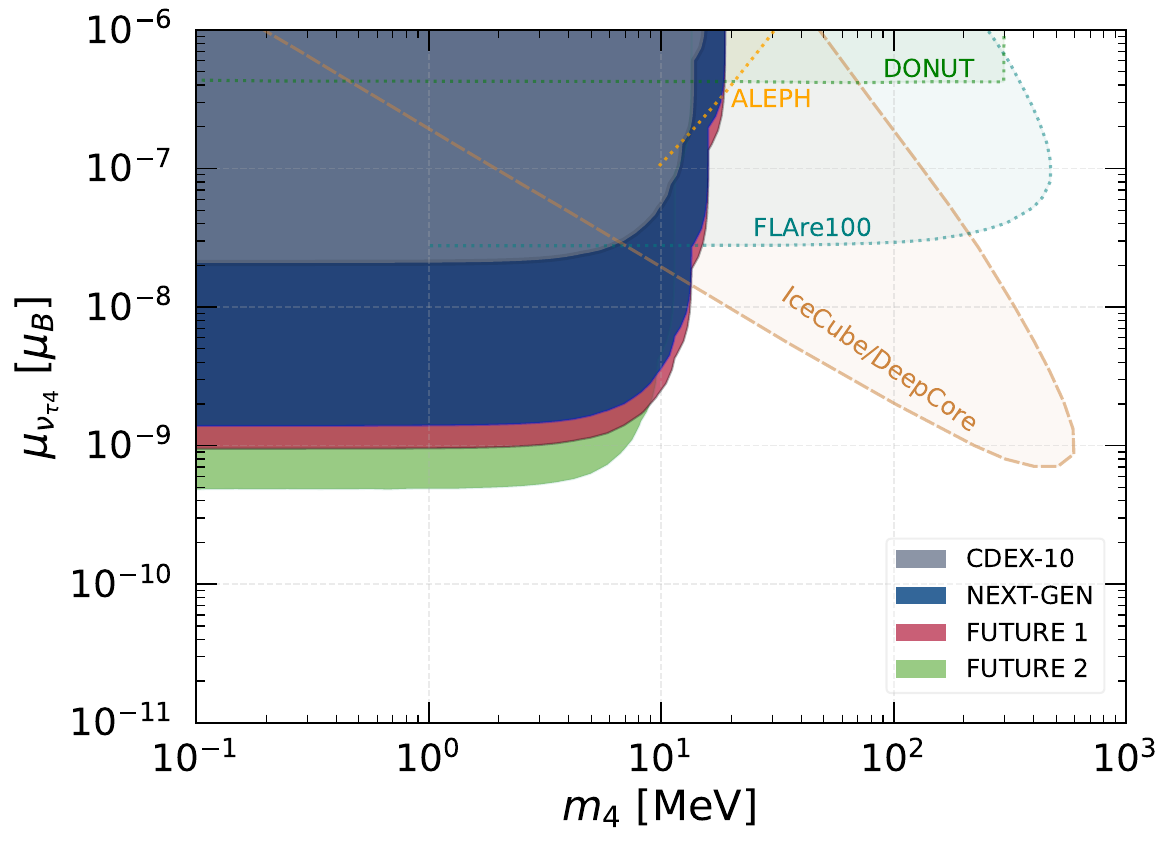}
	\\
	\hspace{10.4mm}	(a) \hspace{80.8mm} (b)
	\caption{(a) 90$\%$ C.L. (2 d.o.f.) exclusion regions on the plane of the active $\nu_\tau$-sterile neutrino transition magnetic moment vs sterile neutrino mass from the current CDEX-10 data, next-generation and future projected scenarios, and (b) comparison with other available experimental constraints (see the text for details).}
	\label{fig:B8hep_flav_tau}
\end{figure*}
\begin{figure*}[ht]
	\centering
	\includegraphics[scale=0.43]{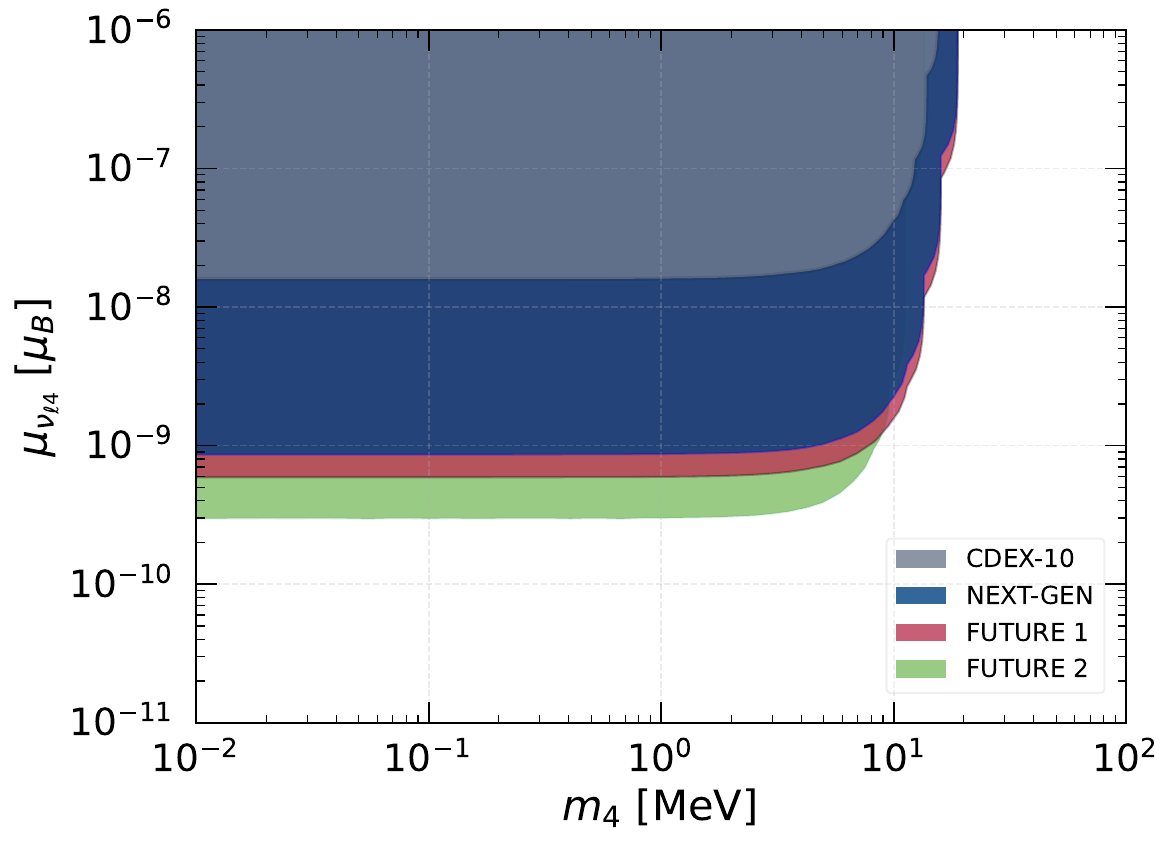}
	\includegraphics[scale=0.43]{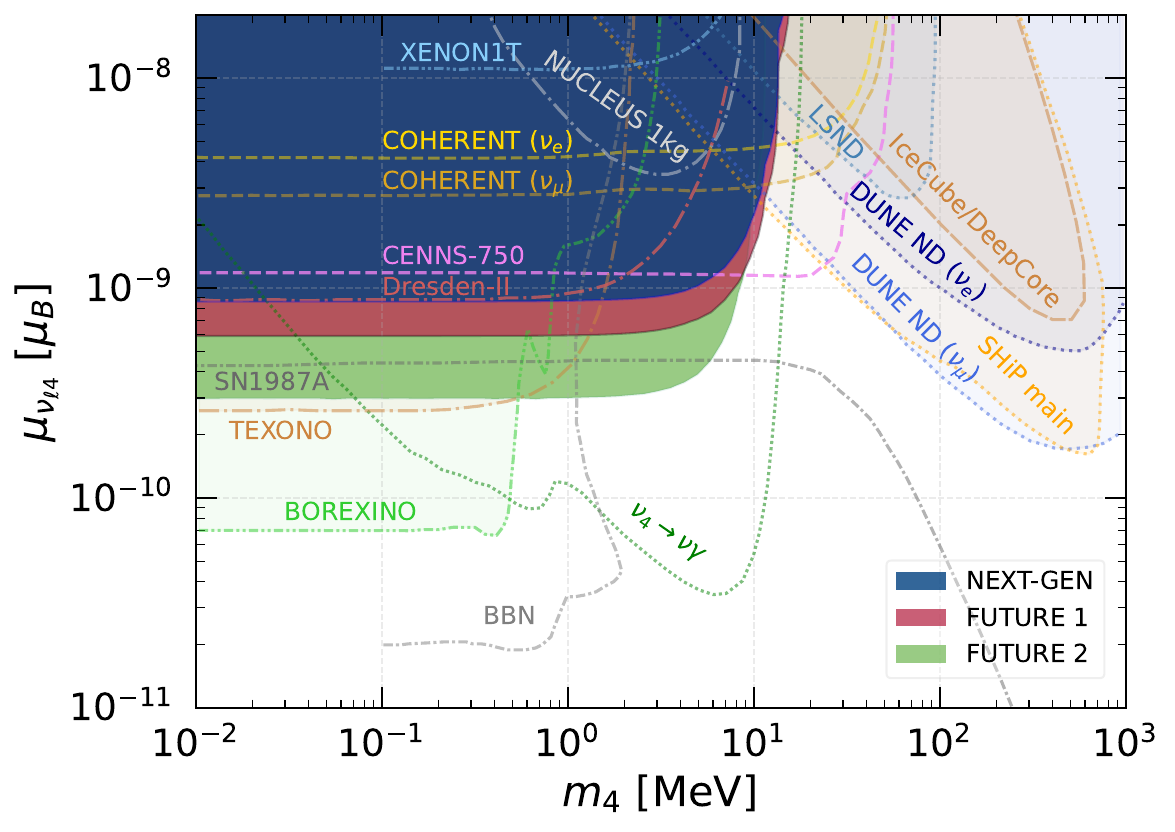}
	\\
	\hspace{10.4mm}	(a) \hspace{80.8mm} (b)
	\caption{(a) 90$\%$ C.L. (2 d.o.f.) exclusion regions on the active flavor-independent $\nu$-sterile neutrino transition magnetic moment vs sterile neutrino mass from the current CDEX-10 data, next-generation and future projected scenarios, and (b) comparison with other available experimental constraints (see the text for details).}
	\label{fig:B8hep_flav_blind}
\end{figure*}

Our analysis results for the active $\nu_{e}$-sterile transition magnetic moment $\mu_{\nu_{e4}}$ is shown in Fig.~\ref{fig:B8hep_flav_e}(a). The upper-limit of $\mu_{\nu_{e 4}}$ derived from CDEX-10 data reaches $2.26 \times 10^{-8} \mu_B$ in the region of $m_4 < 0.1$ MeV. Meanwhile, the next-generation, future 1, and future 2 scenarios provide upper-limits of $\mu_{\nu_{e 4}} \lesssim 1.50 \times 10^{-9}\mu_B$, $\mu_{\nu_{e 4}} \lesssim 1.04 \times 10^{-9}\mu_B$, and $\mu_{\nu_{e 4}} \lesssim 5.08\times 10^{-10}\mu_B$, respectively. 
The future 1 and 2 scenarios yield $30.5 \%$ and $66.2 \%$ more stringent limits than the next-generation scenario, respectively. Compared to the current limit, the next-generation, the future 1 and 2 scenarios yield improvements of $15.1$, $21.7$, and $44.5$ times, respectively.
In Fig.~\ref{fig:B8hep_flav_e}(b), we further compare our projected sensitivities with the previously studied limits relevant to $\mu_{\nu_{e 4}}$.
It can be seen that the projected scenarios yield more stringent constraints than COHERENT CsI+LAr in low-mass regions. 
This can be understood from the lower recoil energy threshold of the projected scenarios, which allows it to more efficiently probe the $1/T_{nr}$ than neutrinos from stopped pion sources. Regarding constraints of neutrino from reactor facilities, the Dresden-II limit is dominated by the next-generation scenario for $m_4 \gtrsim 3$ MeV and is fully covered by the future 2 scenario. Though, in the mass range lower than 1.2 MeV, the projected sensitivities are not as good as the TEXONO limit, they provide better constraints for higher masses.  
As expected, the effect from neutrino-electron scatterings of reactor facilities prevails at this low-mass range compared to CE$\nu$NS at solar neutrino experiments due to the lower energy threshold on the incident neutrinos. Meanwhile, for the expected future facilities, the projected scenarios fully cover the limit of NUCLEUS 1 kg and cover the low-mass region of FLAre100 and CENNS-750 limits. Furthermore, a large portion of the unexplored space of the LSND, DUNE ND, and SHiP main are covered in the MeV mass scale. This is due to the high energy of neutrinos produced in accelerator facilities, hence they are sensitive only at the high-mass region.

Moving on to the muon neutrino case, our results from the current CDEX-10 data, next-generation and two future scenarios are shown in Fig.~\ref{fig:B8hep_flav_mu}(a). The upper limit of the $\mu_{\nu_{\mu 4}}$ from the current data is $2.41 \times 10^{-8} \mu_B$ in the region of $m_4 \lesssim 0.1$ MeV, while the next-generation, future 1, and future 2 scenarios provide $\mu_{\nu_{\mu 4}} \lesssim 1.59 \times 10^{-9}\mu_B$, $\mu_{\nu_{\mu 4}} \lesssim 1.11 \times 10^{-9}\mu_B$, and $\mu_{\nu_{\mu 4}} \lesssim 5.64 \times 10^{-10}\mu_B$, respectively.
The future 1 scenario yields around $30.1\%$ and the future 2 scenario yields around $64.4 \%$ more stringent constraints than the next-generation scenario. On the other hand, the next-generation, the future 1 and 2 scenarios provide improvements of $15.2$, $21.7$, and $42.7$ times, respectively, compared to the current limit. 
We also show the comparison of these results with previous limits in Fig. \ref{fig:B8hep_flav_mu}(b). Our projected sensitivities cover the XENON1T limit and the low-mass regions of COHERENT (CsI+LAr) and CHARM-II, by providing up to two orders of magnitude improvement. 
The projected scenarios also provide improvement to the limits derived from BOREXINO for $m_4\gtrsim 0.6$ MeV, where at low mass-region the limit from the neutrino-electron process still dominates due to low energy threshold. Regarding the expected future facilities, our scenarios reach the low-mass region of FLAre100, CENNS-750, and DUNE ND.
Though our projected constraints are yet to reach the results from DARWIN data for $\lesssim 0.6$ MeV, these give better limits in high mass regimes.
Meanwhile, the limits of NOMAD and the ROI of MiniBooNE dominate in the high-mass region, anticipated from the high neutrino energy of the accelerator facilities.

In Fig.~\ref{fig:B8hep_flav_tau}(a), we show limits derived from the current data and the next-generation, future 1 and future 2 scenarios for the $\tau$ neutrino case. The current upper-limit of the $\mu_{\nu_{\tau 4}}$ is around $2.08\times 10^{-8} \mu_B$ in the region of $m_4 \lesssim 0.1$ MeV, while the next-generation, future 1, and future 2 scenarios provide $\mu_{\nu_{\tau 4}} \lesssim 1.39 \times 10^{-9} \mu_B$, $\mu_{\nu_{\tau 4}} \lesssim 9.50 \times 10^{-10} \mu_B$, and $\mu_{\nu_{\tau 4}} \lesssim 4.86 \times 10^{-10} \mu_B$, respectively. 
The future 1 yields about $31.6 \%$ and future 2 about $65.0 \%$ more stringent limit than the next-generation scenario. Meanwhile, the next-generation, future 1 and 2 scenarios yield $15.0$, $21.9$, and $42.8$ times improvements, respectively,
over the current limit.	
We also present a comparison of our results with previous limits derived from ALEPH, FLAre100, DONUT and IceCube/DC data in Fig.~\ref{fig:B8hep_flav_tau} (b). 
It is seen that the current CDEX-10 data and the projected scenarios yield better sensitivities in the low-mass region than the considered available limits. Particular to the projected scenarios, they improve the limit for DONUT and ALEPH as $m_4 \lesssim 19$ MeV and FLArE-100 as $m_4 \lesssim 17$ MeV. This behavior is anticipated from the low-energy neutrino flux of solar neutrino compared to the neutrino from accelerators in these facilities. Meanwhile, our results cover larger parameter space than the IceCube/DC limit up to $m_4\lesssim 15$ MeV, indicating that solar neutrinos are more sensitive to probe low-mass regions than the atmospheric neutrinos that have slightly larger neutrino energy.

We finally present constraints on the flavor-independent $\nu$-sterile transition magnetic moment $\mu_{\nu_{\ell 4}}$ in Fig.~\ref{fig:B8hep_flav_blind}(a). The upper limit of the $\mu_{\nu_{\ell 4}}$ from the current CDEX-10 data is obtained as $1.59 \times 10^{-8} \mu_B$ for $m_4 \lesssim 0.1$ MeV. Meanwhile, the next-generation, future 1, and future 2 provide upper limits of $\mu_{\nu_{\ell 4}} \lesssim 8.63 \times 10^{-10} \mu_B$, $\mu_{\nu_{\ell 4}} \lesssim 5.91 \times 10^{-10} \mu_B$, and $\mu_{\nu_{\ell 4}} \lesssim 2.97 \times 10^{-10} \mu_B$, respectively. It can be seen that the future 1 scenario yields about $31.5\%$ and the future 2 scenario yields about $65.5 \%$ more stringent limits than the next-generation scenario. Furthermore, over the current limit, the next-generation, future 1 and 2 scenarios provide $18.4$, $26.9$, and $53.3$ times improvements, respectively.	
We provide a comparison of our projected sensitivities of $\mu_{\nu_{\ell 4}}$ with the previous limits in Fig.~\ref{fig:B8hep_flav_blind}(b). 
It is clearly seen that the projected scenarios could provide more robust constraints than some available ones. They fully cover the region of XENON1T, NUCLEUS 1kg and Dresden-II limits.
\begin{table*}[htb]
	\caption{90\% C.L. (2 d.o.f.) upper-limits on the active-sterile neutrino transition magnetic moment. Here, we also include the available limits from COHERENT, FLAre100, and XENON1T in the considered parameter region.
	}
	\begin{center}
	\begin{tabular*}{\textwidth}{@{\extracolsep{\fill}}l c c c c c c c @{}}
							\hline 
							\hline
				\multirow{2}{0.8cm}{TMM $\times 10^{-9} \mu_B$} & \multicolumn{4}{c}{This work} & {COHERENT \cite{DeRomeri:2022twg}} &
				{FLAre100 \cite{Ismail:2021dyp}} & 	
				{XENON1T \cite{Brdar:2020quo}}\\
				\cline{2-5}
				& Current & Next-Gen & Future 1 & Future 2 &  
			\\
				\hline
				$\mu_{\nu_{e 4}}$& $\lesssim 22.6$ &  $ \lesssim 1.50$ & $ \lesssim 1.04$ & $ \lesssim 0.51$ & 
				$ \lesssim 4.20$ &
				$ \lesssim 6.60$ &
				-
				\\
				$\mu_{\nu_{\mu 4}}$ & $ \lesssim 24.1$ & $ \lesssim 1.59$ & $ \lesssim 1.11$ & $ \lesssim 0.56$ &
				$\lesssim 2.77$ &
				$\lesssim 2.25$ &
				$ \lesssim 12.4$ 
				\\
				$\mu_{\nu_{\tau 4}}$ & $ \lesssim 20.8$ & $ \lesssim 1.39$ & $ \lesssim 0.95$ & $ \lesssim 0.49$ & - & $\lesssim 0.28$
				&-
				\\
				$\mu_{\nu_{\ell 4}}$ & $\lesssim 15.9$ &  $ \lesssim 0.86$ & $ \lesssim 0.59$ & $ \lesssim 0.30$ &  - & - & -
				\\
				\hline
			\end{tabular*}
	\end{center}
	\label{tab:sterilemm}
\end{table*}
\begin{figure*}[ht!]
	\centering
	\includegraphics[scale=0.43]{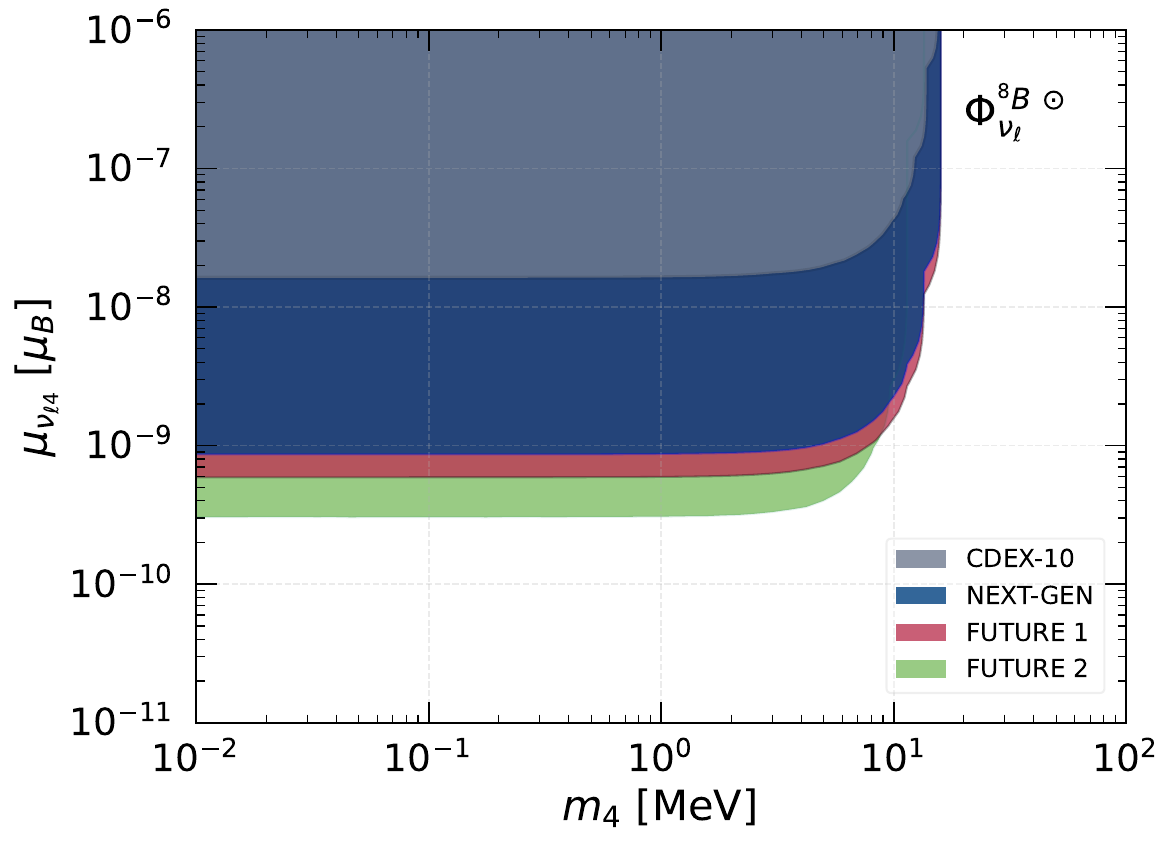}
	\includegraphics[scale=0.43]{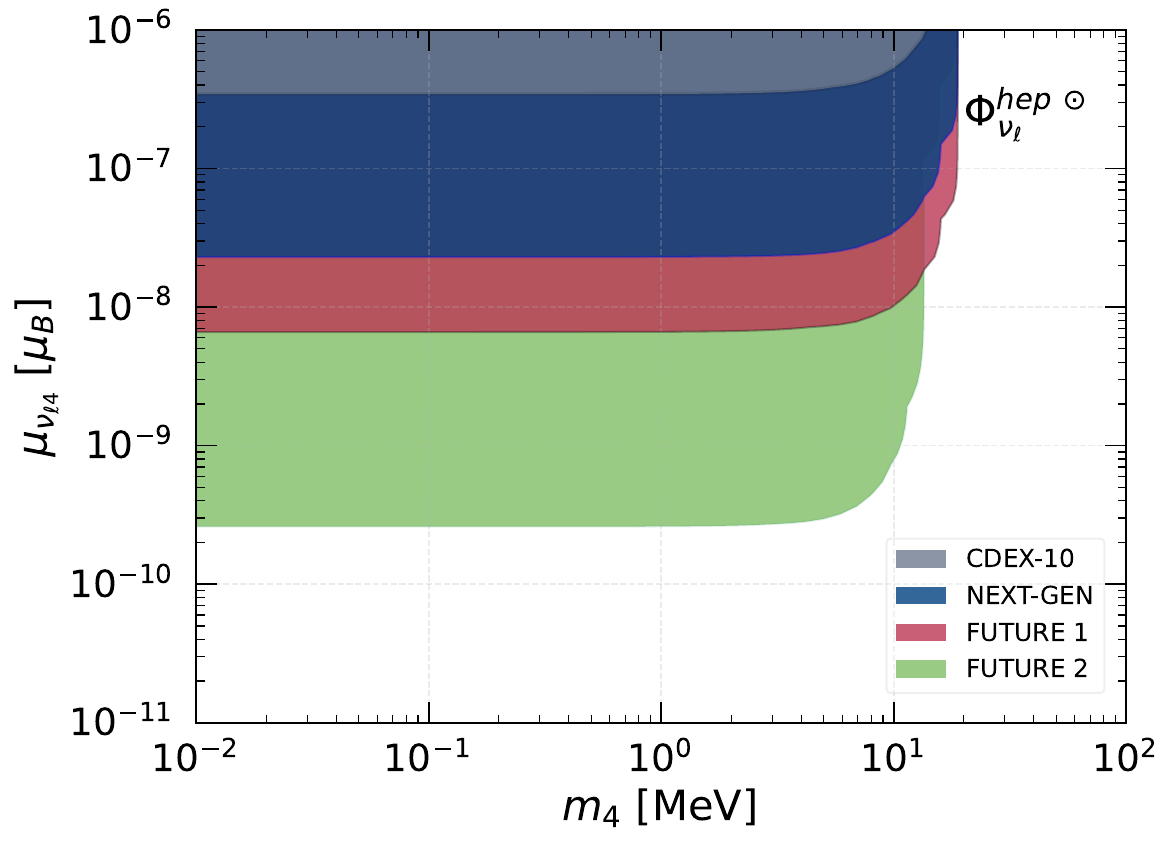}\\
	\hspace{10.4mm}	(a) \hspace{80.8mm} (b)
	\caption{90$\%$ C.L. (2 d.o.f.) exclusion regions on the flavor-independent $\nu$-sterile neutrino transition magnetic moment from the current CDEX-10 data, next-generation and future projected scenarios for (a) the $^8$B and (b) hep solar fluxes, separately.}
	\label{fig:B8_hep_eff}
\end{figure*}
Furthermore, our results reach the previously unexplored regions of the COHERENT ($\nu_e$ and $\nu_\mu$), CENNS-750, as well as LSND, SHiP main, DUNE ND ($\nu_e$ and $\nu_\mu$), and IceCube/DeepCore.  There appear improvements for the region of $m_4\gtrsim0.5$ MeV, which cannot be reached at BOREXINO and TEXONO by measuring neutrino-electron scattering. The CE$\nu$NS process dominates the contributions to the limits in the heavy sterile-mass regime ($m_4 \gtrsim $ MeV) due to kinematic constraints, while the neutrino-electron scattering contribution prevails for a lower energy threshold on the incident neutrinos.
Concerning the cosmological bounds, our results dominate the BBN limit for $m_4 \gtrsim 1$ MeV while the limit of the SN1987A is reached by the future 2 scenario. 
Improvements shown in the projected scenarios to some previous bounds signify the importance of observing low-energy nuclear thresholds and the increase of experimental exposure. Lastly, we show the sensitivity from the $\nu_4 \rightarrow \nu \gamma$ decay to complement our results.

We summarize in Table \ref{tab:sterilemm} the upper-limits derived in this work. Also, we include previously available limits from COHERENT, FLAre100, and XENON1T for comparison. The limits can be directly read from Figs.~\ref{fig:B8hep_flav_e}, \ref{fig:B8hep_flav_mu}, \ref{fig:B8hep_flav_tau}, and \ref{fig:B8hep_flav_blind}. In general, the current data show competitive limits compared to DD experiments such as XENON1T limit, while the projected scenarios provide significant improvements over some previous constraints. 
Lowering the threshold energy and increasing the exposure of experiments in the predicted scenarios improve the limits of the current data by almost two orders of magnitude.
We find that the sensitivities deteriorate at $m_4 \approx 10$ MeV
as the lowest energy solar neutrinos no longer have sufficient energy to create $\nu_4$.

Finally, we also discuss the effects of individual neutrino fluxes from $^8\mathrm{B}$ and $hep$ components on the analysis results in Fig.~\ref{fig:B8_hep_eff}. It can be seen from Fig.~\ref{fig:B8_hep_eff}(a) that the upper-limits from $^8\mathrm{B}$ flux is $\mu_{\nu_{\ell 4}} \lesssim 1.63\times 10^{-8} \mu_B$ for the current data, while $\mu_{\nu_{\ell 4}} \lesssim 8.64\times 10^{-10}\mu_B$, $\mu_{\nu_{\ell 4}} \lesssim 5.91\times 10^{-10}\mu_B$, and $\mu_{\nu_{\ell 4}} \lesssim 3.04\times 10^{-10}\mu_B$ for the next-generation, future 1 and future 2 scenarios, respectively. Meanwhile in Fig.~\ref{fig:B8_hep_eff}(b), the analysis results for ${hep}$ are shown. It is seen that the upper-limit of the current data reaches $\mu_{\nu_{\ell 4}} \lesssim 3.44\times 10^{-7}\mu_B$, while the next-generation reaches $\mu_{\nu_{\ell 4}} \lesssim 2.29\times 10^{-8}\mu_B$, the future 1 reaches $\mu_{\nu_{\ell 4}} \lesssim 6.60\times 10^{-9}\mu_B$, and the future 2 reaches $\mu_{\nu_{\ell 4}} \lesssim 2.60\times 10^{-10}\mu_B$. Consequently, solar neutrino flux of $^8\mathrm{B}$ generally contributes about one order of magnitude more than the one of ${hep}$, originating from the smallness of the latter flux.

\section{Conclusions}\label{sec:conc} 
We have studied the active-sterile neutrino transition magnetic moments through CE$\nu$NS induced by solar neutrinos using CDEX-10 data. The nonzero neutrino magnetic moment is one of the hints suggesting the need for BSM physics in the neutrino sector implied by neutrino mass. 
In the presence of heavy right-handed neutrinos, dipole interactions between the active neutrinos and sterile states face relatively few constraints due to kinematic limits on the production of the new states. 
This phenomenon will be an interesting research topic to be investigated with  the advancement of DD experiments, which also has great potential to detect the solar neutrino CE$\nu$NS. The low-momentum transfer nature of the process is a unique framework to investigate active-sterile transition magnetic moment in the future, given that experimental advancements will be able to set the most stringent terrestrial limits.

We considered both the flavor-independent and flavor-dependent cases of the active-sterile transition magnetic moment. The event-rate spectra are calculated in terms of nuclear recoil energy which can be converted to the electron-equivalent signal using a quenching factor. It can be seen that lighter sterile mass gives rise to a higher expected event rate at low recoil energy regions. We have derived new constraints on the active-sterile neutrino transition magnetic moment from the latest CDEX-10 data. We have suggested three projected scenarios that can be realized following further upgrades of DD experiments which potentially improve the precision of the CE$\nu$NS signal. Accordingly, we then compared our analysis with existing limits from previous works. Our analysis indicates that the projected sensitivities could cover some regions which were previously unexplored for the sterile neutrino mass below $\sim$10 MeV.

The possibility of sterile neutrinos is interesting from a phenomenological standpoint, and we demonstrated the utility of solar neutrinos in the framework of CE$\nu$NS experiments to explore the properties of this idea. 
The low-energy neutrinos from the Sun enable us to derive constraints on the active-sterile neutrino transition magnetic moment and compare them with results from existing facilities.  
We expect more opportunities to examine new physics from future experiments and our results may provide clues for those endeavors. Conclusively, the CE$\nu$NS process remains to be an important tool in searching BSM scenarios with a strong discovery potential.

\section*{Acknowledgments}
We would like to thank Prof. A. B. Balantekin for helpful discussions and suggestions. This work was supported by the Scientific and Technological Research Council of Türkiye (TUBITAK) under the project no: 123F186. A part of the computations reported in this work was performed at the National Academic Network and Information Center (ULAKBIM) of TUBITAK, High Performance and Grid Computing Center (TRUBA Resources). 

\bibliographystyle{unsrt}

\begin{thebibliography}{99}
	\bibitem{Freedman:1973yd} D.~Z.~Freedman, 
	\href{https://doi.org/10.1103/PhysRevD.9.1389}{{Phys. Rev. D} \textbf{9}, 1389 (1974)}.
	
	
	\bibitem{Akimov:2017ade}
	D.~Akimov \textit{et al.} (COHERENT Collaboration),
	\href{https://doi.org/10.1126/science.aao0990}{{Science} \textbf{357}, 1123 (2017)}.
	
	\bibitem{Akimov:2020pdx}
	D.~Akimov \textit{et al.} (COHERENT Collaboration),
	\href{https://doi.org/10.1103/PhysRevLett.126.012002}{{Phys. Rev. Lett.} \textbf{126}, 012002, (2021)}.
	
	\bibitem{COHERENT:2021xmm}
	D.~Akimov \textit{et al.} (COHERENT Collaboration),
	\href{https://doi.org/10.1103/PhysRevLett.129.081801}{{Phys. Rev. Lett.} \textbf{129}, 081801 (2022)}.
	
	\bibitem{Canas2018}	
	B. C. Cañas, E. A. Garcés, O. G. Miranda, and A. Parada,
	\href{https://doi.org/10.1016/j.physletb.2018.07.049} {Phys. Lett. B \textbf{784}, 159 (2018)}.
	
	\bibitem{Cadeddu2019}
	M. Cadeddu and F. Dordei,
	\href{https://doi.org/10.1103/PhysRevD.99.033010}{Phys. Rev. D \textbf{99}, 033010 (2019)}.
	
	
	\bibitem{Harnik:2012ni}
	R.~Harnik, J.~Kopp and P.~A.~N.~Machado,
	\href{http://doi.org/10.1088/1475-7516/2012/07/026}{{JCAP} \textbf{07}, 026 (2012)}.
	
	\bibitem{Ge2018} 
	S.-F. Ge and I. M. Shoemaker, 
	\href{https://doi.org/10.1007/JHEP11(2018)066}{{ J. High Energ. Phys.} \textbf{2018}, 66 (2018)}.
	
	\bibitem{Boehm:2020ltd}
	C.~B\oe{}hm, D.~G.~Cerde\~no, M.~Fairbairn, P.~A.~N.~Machado and A.~C.~Vincent, 
	\href{http://10.1103/PhysRevD.102.115013}{{Phys. Rev. D} \textbf{102}, 115013 (2020)}.
	
	
	
	\bibitem{Schwemberger:2022fjl}
	T.~Schwemberger and T.~T.~Yu,
	\href{https://doi.org/10.1103/PhysRevD.106.015002}{{Phys. Rev. D} \textbf{106}, 015002 (2022)}. 
	
	\bibitem{Mishra:2023jlq}
	N.~Mishra and L.~E.~Strigari,
	\href{https://doi.org/10.1103/PhysRevD.108.063023}{{Phys. Rev. D} \textbf{108}, 063023 (2023)}.
	
	
	\bibitem{Lindner2017}
	M. Lindner, W. Rodejohann, X.-J. Xu, 
	\href{https://doi.org/10.1007/JHEP03(2017)097}{J. High Energy Phys. \textbf{2017}, 97 (2017)}.
	
	
	\bibitem{Liao2017}
	J. Liao and D. Marfatia, 
	\href{https://doi.org/10.1016/j.physletb.2017.10.046} {Phys. Lett. B \textbf{775}, 54 (2017)}.
	
	
	\bibitem{Giunti:2019xpr}
	C.~Giunti,
	\href{https://doi.org/10.1103/PhysRevD.101.035039}{{Phys. Rev. D} \textbf{101}, 035039 (2020)}. 
	
	\bibitem{Mustamin:2021mtq}
	M.~F.~Mustamin and M.~Demirci,
	\href{https://doi.org/10.1007/s13538-021-00867-x}{{Braz. J. Phys.} \textbf{51}, 813 (2021)}.
	
	\bibitem{Chatterjee:2023} S. S. Chatterjee, S. Lavignac, O. G. Miranda, and G. Sanchez Garcia, 
	\href{https://link.aps.org/doi/10.1103/PhysRevD.107.055019}{{Phys. Rev. D} \textbf{107}, 055019 (2023)}.	
	
	\bibitem{Dent:2016wcr}
	J.~B.~Dent, B.~Dutta, S.~Liao, J.~L.~Newstead, L.~E.~Strigari and J.~W.~Walker,
	\href{https://doi.org/10.1103/PhysRevD.96.095007}{{Phys. Rev. D} \textbf{96}, 095007 (2017)}. 
	
	
	\bibitem{Barranco:2011wx}
	J.~Barranco, A.~Bolanos, E.~A.~Garces, O.~G.~Miranda and T.~I.~Rashba,
	\href{https://doi.org/10.1142/S0217751X12501473}{{Int. J. Mod. Phys. A} \textbf{27}, 1250147 (2012)}. 
	
	\bibitem{AristizabalSierra:2018eqm}
	D.~Aristizabal Sierra, V.~De Romeri and N.~Rojas,
	\href{https://doi.org/10.1103/PhysRevD.98.075018}{{Phys. Rev. D} \textbf{98}, 075018 (2018)}. 
	
	
	\bibitem{Flores:2021kzl}
	L.~J.~Flores, N.~Nath and E.~Peinado,
	\href{https://doi.org/10.1103/PhysRevD.105.055010}{Phys. Rev. D \textbf{105}, 055010 (2022)}. 
	
	
	\bibitem{Farzan:2018gtr}
	Y.~Farzan, M.~Lindner, W.~Rodejohann and X.~J.~Xu,
	\href{https://doi.org/10.1007/JHEP05(2018)066}{{J. High Energ. Phys}  \textbf{2018}, 066 (2018)}.
	
	
	\bibitem{Demirci:2021zci}
	M.~Demirci and M.~F.~Mustamin,
	\href{https://doi.org/10.31526/ACP.BSM-2021.31}{{Andromeda Proceedings}, BSM21 (2021)}.
	
	\bibitem{Cadeddu:2020nbr}
	M.~Cadeddu, N.~Cargioli, F.~Dordei, C.~Giunti, Y.~F.~Li, E.~Picciau and Y.~Y.~Zhang,
	\href{http://doi.org/10.1007/JHEP01(2021)116}{{J. High Energ. Phys} \textbf{2021}, 116 (2021)}.
	
	\bibitem{AtzoriCorona:2022moj}
	M.~Atzori Corona, M.~Cadeddu, N.~Cargioli, F.~Dordei, C.~Giunti, Y.~F.~Li, E.~Picciau, C.~A.~Ternes and Y.~Y.~Zhang,
	\href{http://doi.org/10.1007/JHEP05(2022)109}{{JHEP} \textbf{2022}, 109 (2022)}.
	
	\bibitem{Demirci:2024}
	M.~Demirci and M.~F.~Mustamin, 
	\href{https://doi.org/10.1103/PhysRevD.109.015021}{{Phys. Rev. D} \textbf{109}, 015021 (2024)}.
	
	\bibitem{DeRomeri:2024dbv}
	V.~De Romeri, D.~K.~Papoulias and C.~A.~Ternes,
	\href{https://doi.org/10.1007/JHEP05(2024)165}{JHEP \textbf{05}, 165 (2024)}.
	
	\bibitem{Herrera:2023xun}
	G.~Herrera,
	\href{https://doi.org/10.1007/JHEP05(2024)288}{JHEP \textbf{05}, 288 (2024)}.
	
	\bibitem{Giunti:2015} C. Giunti and A. Studenikin, 
	\href{https://doi.org/10.1103/RevModPhys.87.531}{Rev. Mod. Phys. \textbf{87}, 531 (2015)}.
	
	\bibitem{Studenikin2020} A. I. Studenikin and K. A. Kouzakov, 
	\href{https://doi.org/10.3103/S0027134920050215}{Moscow University Physics Bulletin \textbf{75}, 379–397 (2020)}.
	
	\bibitem{Giunti2023}
	C. Giunti and C. A. Ternes, 
	\href{https://doi.org/10.1103/PhysRevD.108.095044}{{Phys. Rev. D} \textbf{108}, 095044 (2023)}.
	
	
	\bibitem{AtzoriCorona:2022} 	M.~Atzori Corona, \textit{et al.}, 
	\href{https://doi.org/10.1007/JHEP09(2022)164}{{J. High Energ. Phys} \textbf{2022}, 164 (2022)}.
	
	\bibitem{Khan:2023}
	A. N.~Khan,
	\href{https://doi.org/10.1016/j.nuclphysb.2022.116064}{Nucl. Phys. B \textbf{986}, 116064 (2023)}.
	
	
	\bibitem{Khan:2023b}
	A. N.~Khan, 
	\href{https://doi.org/10.1016/j.physletb.2022.137650}{Phys. Lett. B \textbf{837}, 137650 (2023)}.
	
	\bibitem{Co:2020gwl}
	G.~Co', M.~Anguiano and A.~M.~Lallena,
	\href{https://doi.org/10.1088/1475-7516/2020/04/044}{JCAP \textbf{04}, 044 (2020)}.
	
	\bibitem{Coloma:2020nhf}
	P.~Coloma, I.~Esteban, M.~C.~Gonzalez-Garcia and J.~Menendez,
	\href{https://doi.org/10.1007/JHEP08(2020)030}{JHEP \textbf{08}, 030 (2020)}.
	
	\bibitem{Super-Kamiokande:1998kpq}
	Y.~Fukuda \textit{et al.} (Super-Kamiokande Collaboration),
	\href{https://doi.org/10.1103/PhysRevLett.81.1562}{Phys. Rev. Lett. \textbf{81}, 1562-1567 (1998)}.
	
	\bibitem{SNO:2001kpb}
	Q.~R.~Ahmad \textit{et al.} (SNO Collaboration),
	\href{https://doi.org/10.1103/PhysRevLett.87.071301}{Phys. Rev. Lett. \textbf{87}, 071301 (2001)}.
	
	\bibitem{SNO:2002tuh}
	Q.~R.~Ahmad \textit{et al.} (SNO Collaboration),
	\href{https://doi.org/10.1103/PhysRevLett.89.011301}{Phys. Rev. Lett. \textbf{89}, 011301 (2002)}.
	
	\bibitem{Pontecorvo:1967fh}
	B.~Pontecorvo,
	\href{http://www.jetp.ras.ru/cgi-bin/e/index/e/26/5/p984?a=list}{Zh. Eksp. Teor. Fiz. \textbf{53}, 1717-1725 (1967)}.
	
	\bibitem{Kusenko:2009up}
	A.~Kusenko,
	\href{https://doi.org/10.1016/j.physrep.2009.07.004}{Phys. Rept. \textbf{481}, 1-28 (2009)}.
	
	\bibitem{Dasgupta:2021ies}
	B.~Dasgupta and J.~Kopp,
	\href{https://doi.org/10.1016/j.physrep.2021.06.002}{Phys. Rept. \textbf{928}, 1-63 (2021)}.
	
	\bibitem{MiniBooNE:2010idf}
	A.~A.~Aguilar-Arevalo \textit{et al.} (MiniBooNE Collaboration),
	\href{https://doi.org/10.1103/PhysRevLett.105.181801}{Phys. Rev. Lett. \textbf{105}, 181801 (2010)}.
	
	\bibitem{Arguelles:2021meu}
	C.~A.~Arg\"uelles , I.~Esteban, M.~Hostert, K.~J.~Kelly, J.~Kopp, P.~A.~N.~Machado, I.~Martinez-Soler and Y.~F.~Perez-Gonzalez,
	\href{https://doi.org/10.1103/PhysRevLett.128.241802}{Phys. Rev. Lett. \textbf{128}, 241802 (2022)}.
	
	\bibitem{LSND:1997vun}
	C.~Athanassopoulos \textit{et al.} (LSND Collaboration),
	\href{https://doi.org/10.1103/PhysRevLett.81.1774}{Phys. Rev. Lett. \textbf{81}, 1774-1777 (1998)}.
	
	\bibitem{McLaughlin:1999pd}
	G.~C.~McLaughlin, J.~M.~Fetter, A.~B.~Balantekin and G.~M.~Fuller,
	\href{https://doi.org/10.1103/PhysRevC.59.2873}{Phys. Rev. C \textbf{59}, 2873-2887 (1999)}.
	
	\bibitem{Xiong:2019nvw}
	Z.~Xiong, M.~R.~Wu and Y.~Z.~Qian,
	\href{https://doi.org/10.3847/1538-4357/ab2870}{The Astrophysical Journal \textbf{880}, 2 (2019)}.
	
	\bibitem{Dodelson:1993je}
	S.~Dodelson and L.~M.~Widrow,
	Sterile-neutrinos as dark matter,
	\href{https://doi.org/10.1103/PhysRevLett.72.17}{Phys. Rev. Lett. \textbf{72}, 17-20 (1994)}.
	
	\bibitem{Balantekin2014} A. B. Balantekin and N. Vassh, 
	\href{https://link.aps.org/doi/10.1103/PhysRevD.89.073013}{{Phys. Rev. D}~\textbf{89}, 073013 (2014)}.
	
	\bibitem{Khan:2022bcl}
	A.~N.~Khan,
	\href{https://doi.org/10.1007/JHEP01(2023)052}{JHEP \textbf{01}, 052 (2023)}.
	
	\bibitem{Mirizzi:2012we}
	A.~Mirizzi, N.~Saviano, G.~Miele and P.~D.~Serpico,
	\href{https://doi.org/10.1103/PhysRevD.86.053009}{{Phys. Rev. D} \textbf{86}, 053009 (2012)}.
	
	\bibitem{Domokos:1996cn}
	G.~Domokos and S.~Kovesi-Domokos,
	\href{https://10.1103/PhysRevD.55.R2526}{{Phys. Rev. D} \textbf{55}, 2526-2529 (1997)}.
	
	\bibitem{Gninenko:1998nn}
	S.~N.~Gninenko and N.~V.~Krasnikov,
	\href{https://doi.org/10.1016/S0370-2693(99)00130-6}{{Phys. Lett. B} \textbf{450}, 165 (1999)}.
	
	\bibitem{Primakoff:1951iae}
	H.~Primakoff,
	\href{https://doi.org/10.1103/PhysRev.81.899}{{Phys. Rev.} \textbf{81}, 899 (1951)}.
	
	\bibitem{McKeen2010} D. McKeen, and M. Pospelov, 
	\href{https://link.aps.org/doi/10.1103/PhysRevD.82.113018}{{Phys. Rev. D}~\textbf{82}, 113018 (2010)}.
	
	\bibitem{Magill:2018jla}
	G.~Magill, R.~Plestid, M.~Pospelov and Y.~D.~Tsai,
	\href{https://doi.org/10.1103/PhysRevD.98.115015}{{Phys. Rev. D} \textbf{98}, 115015 (2018)}.
	
	\bibitem{Blanco2020} C. Blanco, D. Hooper, and P. Machado, 
	\href{https://link.aps.org/doi/10.1103/PhysRevD.101.075051}{{Phys. Rev. D} \textbf{101}, 075051 (2020)}.
	
	\bibitem{Schwetz2021}
	T. Schwetz, A. Zhoua and Jing-Yu Zhua, 
	\href{https://doi.org/10.1007/JHEP07(2021)200}{J. High Energ. Phys \textbf{2021}, 200 (2021)}.
	
	\bibitem{Gninenko2011} 
	S. N. Gninenko,
	\href{https://link.aps.org/doi/10.1103/PhysRevD.83.015015}{{Phys. Rev. D} \textbf{83}, 015015 (2011)}.
	
	\bibitem{DeRomeri:2022twg}	
	V.~De Romeri, O.~G.~Miranda, D.~K.~Papoulias, G.~Sanchez Garcia, M.~T\'ortola and J.~W.~F.~Valle,
	\href{http://doi.org/10.1007/JHEP04(2023)035}{{ J. High Energ. Phys.} \textbf{2023}, 035 (2023)}.
	
	\bibitem{Bolton2022} 
	P. D. Bolton, F. F. Deppisch, K. Fridell, J. Harz, C. Hati, and S. Kulkarni,
	\href{https://doi.org/10.1103/PhysRevD.106.035036}{{Phys. Rev. D} \textbf{106}, 035036 (2022)}.
	
	\bibitem{Ismail:2021dyp}
	A.~Ismail, S.~Jana and R.~Mammen~Abraham,
	\href{https://doi.org/10.1103/PhysRevD.105.055008}{{Phys. Rev. D} \textbf{105}, 055008 (2022)}.
	
	\bibitem{Coloma:2017ppo}
	P.~Coloma, P.~A.~N.~Machado, I.~Martinez-Soler and I.~M.~Shoemaker,
	\href{https://doi.org/10.1103/PhysRevLett.119.201804}{{Phys. Rev. Lett.} \textbf{119}, 201804 (2017)}.
	
	\bibitem{Plestid:2020vqf}
	R.~Plestid,
	\href{https://doi.org/10.1103/PhysRevD.104.075027}{{Phys. Rev. D} \textbf{104}, 075027 (2021)}.
	
	\bibitem{Atkinson:2021rnp}
	M.~Atkinson, P.~Coloma, I.~Martinez-Soler, N.~Rocco, and I.~M.~Shoemaker,
	\href{https://doi.org/10.1007/JHEP04(2022)174}{{J. High Energ. Phys.} \textbf{2022}, 174 (2022)}.
	
	\bibitem{Miranda2021}
	O. G. Miranda, D. K. Papoulias, O. Sanders, M. Tortola and J. W. F. Valle, 
	\href{https://doi.org/10.1007/JHEP12(2021)191}{{J. High Energ. Phys} \textbf{2021}, 191 (2021)}.
	
	\bibitem{Li2022} Y.-F. Li, and S. Y. Xia, 
	\href{https://doi.org/10.1103/PhysRevD.106.095022}{{Phys. Rev. D} \textbf{106}, 095022 (2022)}.
	
	\bibitem{Brdar:2020quo}
	V.~Brdar, A.~Greljo, J.~Kopp and T.~Opferkuch,
	\href{https://doi.org/10.1088/1475-7516/2021/01/039}{{JCAP} \textbf{01}, 039 (2021)}.
	
	\bibitem{Huang:2022pce}
	G.~y.~Huang, S.~Jana, M.~Lindner and W.~Rodejohann,
	\href{https://doi.org/10.1016/j.physletb.2023.137842}{Phys. Lett. B \textbf{840}, 137842 (2023)}.
	
	\bibitem{Antusch:2016ejd}
	S.~Antusch, E.~Cazzato and O.~Fischer,
	\href{https://doi.org/10.1142/S0217751X17500786}{Int. J. Mod. Phys. A \textbf{32}, 1750078 (2017)}.
	
	\bibitem{Balantekin:2023jlg}
	A.~B.~Balantekin, G.~M.~Fuller, A.~Ray and A.~M.~Suliga,
	\href{https://doi.org/10.1103/PhysRevD.108.123011}{Phys. Rev. D \textbf{108}, 123011 (2023)}.
	
	\bibitem{CDEX:2022mlp}	X.~P.~Geng \textit{et al.} (CDEX Collaboration),
	\href{https://doi.org/10.1103/PhysRevD.107.112002}{{Phys. Rev. D}~\textbf{107}, 112002 (2023)}.
	
	
	\bibitem{CDEX:2018lau} H.~Jiang \textit{et al.} (CDEX Collaboration),
	\href{http://doi.org/10.1103/PhysRevLett.120.241301}{{Phys. Rev. Lett.} \textbf{120}, 241301 (2018)}. 
	
	\bibitem{XENON:2020rca}
	E.~Aprile \textit{et al.} (XENON Collaboration),
	\href{https://doi.org/10.1103/PhysRevD.102.072004}{Phys. Rev. D \textbf{102}, 072004 (2020)}.
	
	\bibitem{DUNE:2020ypp}
	B.~Abi \textit{et al.} (DUNE Collaboration),
	\href{https://lss.fnal.gov/archive/2020/pub/fermilab-pub-20-025-nd.pdf}{FERMILAB-PUB-20-025-ND, FERMILAB-DESIGN-2020-02},
	\href{https://doi.org/10.48550/arXiv.2002.03005}{arXiv:2002.03005 [hep-ex]}
	
	\bibitem{Borexino:2007kvk} C.~Arpesella \textit{et al.} (Borexino Collaboration),
	\href{https://doi.org/10.1016/j.physletb.2007.09.054}{{Phys. Lett. B} \textbf{658}, 101-108 (2008)}.
	
	\bibitem{TEXONO:2009knm}
	M.~Deniz \textit{et al.} (TEXONO Collaboration,
	\href{https://doi.org/10.1103/PhysRevD.81.072001}{{Phys. Rev. D} \textbf{81}, 072001 (2010)}.
	
	\bibitem{Colaresi:2022obx}
	J.~Colaresi, J.~I.~Collar, T.~W.~Hossbach, C.~M.~Lewis and K.~M.~Yocum,
	\href{https://doi.org/10.1103/PhysRevLett.129.211802}{Phys. Rev. Lett. \textbf{129}, 211802 (2022)}.
	
	\bibitem{NUCLEUS:2019kxv}
	J.~Rothe \textit{et al.} (NUCLEUS Collaboration),
	\href{https://doi.org/10.1007/s10909-019-02283-7}{J. Low Temp. Phys. \textbf{199}, 433-440 (2019)}.
	
	\bibitem{CHARM:1983ayi}
	F.~Bergsma \textit{et al.} (CHARM Collaboration),
	\href{https://doi.org/10.1016/0370-2693(83)90275-7}{Phys. Lett. B \textbf{128}, 361 (1983)}.
	
	\bibitem{DONUT:2001zvi}
	R.~Schwienhorst \textit{et al.} (DONUT Collaboration),
	\href{https://doi.org/10.1016/S0370-2693(01)00746-8}{Phys. Lett. B \textbf{513}, 23-29 (2001)}.
	
	\bibitem{NOMAD:1997pcg}
	J.~Altegoer \textit{et al.} (NOMAD Collaboration),
	\href{https://doi.org/10.1016/S0168-9002(97)01079-6}{Nucl. Instrum. Meth. A \textbf{404}, 96-128 (1998)}.
	
	\bibitem{ALEPH:1991qhf}
	D.~Decamp \textit{et al.} (ALEPH Collaboration),
	\href{https://doi.org/10.1016/0370-1573(92)90177-2}{Phys. Rept. \textbf{216}, 253-340 (1992)}.
	
	\bibitem{IceCube:2015vkp}
	M.~G.~Aartsen \textit{et al.} (IceCube Collaboration),
	\href{https://doi.org/10.1103/PhysRevD.93.022001}{Phys. Rev. D \textbf{93}, 022001 (2016)}.
	
	\bibitem{COHERENT:2019kwz}
	D.~Akimov \textit{et al.} (COHERENT Collaboration),
	\href{https://doi.org/10.1103/PhysRevD.102.052007}{Phys. Rev. D \textbf{102}, 052007 (2020)}.
	
	\bibitem{Batell:2021blf}
	B.~Batell, J.~L.~Feng and S.~Trojanowski,
	\href{https://doi.org/10.1103/PhysRevD.103.075023}{Phys. Rev. D \textbf{103}, 075023 (2021)}.
	
	\bibitem{Alekhin:2015byh}
	S.~Alekhin, \textit{et al.} 
	\href{https://doi.org/10.1088/0034-4885/79/12/124201}
	{Rept. Prog. Phys. \textbf{79}, 124201 (2016)}.
	
	\bibitem{DARWIN:2016hyl}
	J.~Aalbers \textit{et al.} (DARWIN Collaboration),
	\href{https://doi.org/10.1088/1475-7516/2016/11/017}{JCAP \textbf{11}, 017 (2016)}.
	
	
	\bibitem{PDG2020} R. L. Workman et al. (Particle Data Group), Review of Particle Physics, 
	\href{https://doi.org/10.1093/ptep/ptac097}{Prog. Theor. Exp. Phys. \textbf{2022}, 083C01 (2022)}.
	
	
	\bibitem{Klein1999} S. R. Klein and J. Nystrand, 
	\href{https://doi.org/10.1103/PhysRevC.60.014903}{{Phys. Rev. C} \textbf{60}, 014903 (1999)}.
	
	
	\bibitem{Vogel:1989iv}
	P.~Vogel and J.~Engel,
	\href{https://doi.org/10.1103/PhysRevD.39.3378}{Phys. Rev. D \textbf{39}, 3378 (1989)}.
	
	\bibitem{Bahcall:2004mq}
	J.~N.~Bahcall and A.~M.~Serenelli,
	\href{http://doi.org/10.1086/429883}{{Astrophys. J} \textbf{626}, 530 (2005)}.
	
	\bibitem{Bahcall:2004pz}
	J.~N.~Bahcall, A.~M.~Serenelli and S.~Basu,
	\href{http://doi.org/10.1086/428929}{{Astrophys. J. Lett.} \textbf{621}, L85-L88 (2005)}.
	
	\bibitem{Maltoni:2015kca}
	M.~Maltoni and A.~Y.~Smirnov,
	\href{http://doi.org/10.1140/epja/i2016-16087-0}{{Eur. Phys. J. A} \textbf{52}, 87 (2016)}.

	
	\bibitem{Esteban:2020cvm}
	I.~Esteban, M.~C. Gonzalez-Garcia, M.~Maltoni, T.~Schwetz, and A.~Zhou, 
	\href{http://dx.doi.org/10.1007/JHEP09(2020)178}{{J. High Energ. Phys}  \textbf{2020}, 178 (2020)}.
	
	
	\bibitem{Lindhard:1963}
	J. Lindhard, V. Nielsen, M. Scharff, and P. V. Thomsen, 
	\href{https://www.osti.gov/biblio/4701226}{Kgl. Danske Videnskab., Selskab. Mat. Fys. Medd. 33, 1, (1963)}.
	
	\bibitem{Bonhomme:2022lcz}
	A.~Bonhomme, \textit{et al.},
	\href{https://doi.org/10.1140/epjc/s10052-022-10768-1}{{Eur. Phys. J. C}  \textbf{82}, 815 (2022)}.
	
	\bibitem{Essig:2018tss}
	R.~Essig, M.~Sholapurkar and T.~T.~Yu,
	\href{https://doi.org/10.1103/PhysRevD.97.095029}{{Phys. Rev. D}  \textbf{97}, 095029 (2018)}.
	
	\bibitem{Collar:2021fcl}
	J.~I.~Collar, A.~R.~L.~Kavner and C.~M.~Lewis,
	\href{https://doi.org/10.1103/PhysRevD.103.122003}{{Phys. Rev. D} \textbf{103} 122003, (2021)}.
	
	\bibitem{Fogli:2002pt}
	G.~L.~Fogli, E.~Lisi, A.~Marrone, D.~Montanino and A.~Palazzo,
	\href{http://doi.org/10.1103/PhysRevD.66.053010}{{Phys. Rev. D} \textbf{66}, 053010 (2002)}.
	
	\bibitem{PandaX:2018wtu}
	H.~Zhang \textit{et al.} (PandaX Collaboration),
	\href{http://doi.org/10.1007/s11433-018-9259-0}{{Sci. China Phys. Mech. Astron.}  \textbf{62}, 31011 (2019)}.
	
	\bibitem{PandaX:2024muv}
	Z.~Bo \textit{et al.} (PandaX Collaboration),
	\href{https://arxiv.org/abs/2407.10892}{arXiv:2407.10892 [hep-ex]}.
	
	\bibitem{XENON:2020kmp} 
	E.~Aprile \textit{et al.} (XENON Collaboration),
	\href{http://doi.org/10.1088/1475-7516/2020/11/031}{{JCAP} \textbf{11}, 031 (2020)}. 
	
	\bibitem{XENON:2024} 
	E.~Aprile \textit{et al.} (XENON Collaboration),
	\href{https://doi.org/10.48550/arXiv.2408.02877}{arXiv:2408.02877 [nucl-ex]}.
	
	
	\bibitem{EDELWEISS:2022ktt}
	E.~Armengaud \textit{et al.} (EDELWEISS Collaboration),
	\href{http://doi.org/10.1103/PhysRevD.106.062004}{{Phys. Rev. D} \textbf{106}, 062004 (2022)}.
	
	\bibitem{SuperCDMS:2016wui}
	R.~Agnese \textit{et al.} (SuperCDMS Collaboration),
	\href{http://doi.org/10.1103/PhysRevD.95.082002}{{Phys. Rev. D} \textbf{95}, 082002 (2017)}.
	
	\bibitem{SENSEI:2020dpa}
	L.~Barak \textit{et al.} (SENSEI Collaboration),
	\href{http://doi.org/10.1103/PhysRevLett.125.171802}{{Phys. Rev. Lett.} \textbf{125}, 171802, (2020)}.
	
	\bibitem{Geng:2023yei}
	X.~P.~Geng, \textit{et al.} (CDEX Collaboration),
		\href{http://doi.org/10.1088/1475-7516/2024/07/009}{{JCAP} \textbf{2024}, 009, (2024)}.
	
	\bibitem{CDEX:2013kpt}
	K.~J.~Kang \textit{et al.} (CDEX Collaboration),
	\href{https://doi.org/10.1007/s11467-013-0349-1}{Front. Phys. (Beijing) \textbf{8} 412-437, (2013)}.
	
	
\end{thebibliography}

\end{document}